\documentclass[conference]{IEEEtran}
\usepackage{fancyhdr}
\usepackage[utf8]{inputenc} \usepackage[T1]{fontenc}

\usepackage{microtype}
\usepackage[bookmarksnumbered,unicode]{hyperref}
\usepackage{graphicx}
\usepackage{textcomp}
\usepackage{cite}
\usepackage{amsmath,amssymb,amsfonts}

\usepackage{listings}
\usepackage[font=small]{caption}
\usepackage{subfig}
\usepackage{grffile} 

\renewcommand{\figurename}{Figure}
\renewcommand{\tablename}{Table}
\renewcommand{\thetable}{\arabic{table}}

\hypersetup{hidelinks}
\hypersetup{pdflang={English},
    pdfdisplaydoctitle}

\usepackage{footnote}
\makesavenoteenv{table}

\newcommand{\zap}[1]{}

\lstdefinelanguage{Chapel}
  {morekeywords={coforall,forall},
  sensitive=true,
  morecomment=[l]{//},
  morecomment=[s]{/*}{*/},
  morestring=[b]",
  morestring=[b]',
}

\lstdefinelanguage{X10}
  {morekeywords={},
  sensitive=true,
  morecomment=[l]{//},
  morecomment=[s]{/*}{*/},
  morestring=[b]",
  morestring=[b]',
}

\lstdefinestyle{codeblock}{
  basicstyle=\ttfamily\footnotesize\linespread{0.75}\selectfont,
}

\lstdefinestyle{codeblocktop}{
  basicstyle=\ttfamily\footnotesize\linespread{0.75}\selectfont,
  float=t!,
  floatplacement=t!,
}
 
\begin{document}

\title{Task Bench: A Parameterized Benchmark for Evaluating Parallel Runtime Performance}

\newcommand{\slac} {\IEEEauthorrefmark{1}}
\newcommand{\lanl} {\IEEEauthorrefmark{2}}
\newcommand{\purdue} {\IEEEauthorrefmark{3}}
\newcommand{\stanford} {\IEEEauthorrefmark{4}}
\newcommand{\utk} {\IEEEauthorrefmark{5}}
\newcommand{\nvidia} {\IEEEauthorrefmark{6}}
\newcommand{\correspondingauthor} {$^1$}
\newcommand{\mccormick} {M\raise .65ex\hbox{c}Cormick}

\author{
\IEEEauthorblockN{Elliott~Slaughter\slac\correspondingauthor, Wei~Wu\lanl\correspondingauthor, Yuankun~Fu\purdue, Legend~Brandenburg\stanford, Nicolai~Garcia\stanford, Wilhem~Kautz\stanford, \\
Emily~Marx\stanford, Kaleb~S.~Morris\stanford, Qinglei~Cao\utk, George~Bosilca\utk, Seema~Mirchandaney\slac, \\
Wonchan~Lee\nvidia, Sean~Treichler\nvidia, Patrick~\mccormick\lanl, Alex~Aiken\stanford}
\IEEEauthorblockA{\slac{}SLAC~National~Accelerator~Laboratory, \lanl{}Los~Alamos~National~Laboratory, \purdue{}Purdue~University,\\
\stanford{}Stanford~University, \utk{}University~of~Tennessee,~Knoxville, \nvidia{}NVIDIA \\
\correspondingauthor{}Corresponding authors: \url{eslaught@slac.stanford.edu}, \url{wwu@lanl.gov}}
}

\maketitle

\thispagestyle{fancy}
\lhead{}
\rhead{}
\chead{}
\lfoot{\footnotesize{
SC20, November 9-19, 2020, Is Everywhere We Are
\newline 978-1-7281-9998-6/20/\$31.00 \copyright 2020 IEEE}}
\rfoot{}
\cfoot{}
\renewcommand{\headrulewidth}{0pt}
\renewcommand{\footrulewidth}{0pt}

\begin{abstract}
We present Task Bench, a \emph{parameterized} benchmark designed to
explore the performance of distributed programming systems under a
variety of application scenarios. Task Bench dramatically lowers the
barrier to benchmarking and comparing multiple programming systems by
making the implementation for a given system orthogonal to the
benchmarks themselves: every benchmark constructed with Task Bench
runs on every Task Bench implementation. Furthermore, Task Bench's
parameterization enables a wide variety of benchmark scenarios that
distill the key characteristics of larger applications.

To assess the effectiveness and overheads of the tested systems, we
introduce a novel metric, \emph{minimum effective task granularity}
(METG). We conduct a comprehensive study with 15 programming systems
on up to 256 Haswell nodes of the Cori supercomputer. Running at
scale, 100\textmu{}s-long tasks are the finest granularity that any
system runs efficiently with current technologies. We also study each
system's scalability, ability to hide communication and mitigate load
imbalance.
\end{abstract}
 
\section{Introduction}
\label{sec:introduction}

The challenge of parallel and distributed computation has led to a
wide variety of proposals for programming models, languages, and
runtime systems. While these systems are well-represented in the literature, comprehensive and comparative performance evaluations
remain difficult to find. Our
goal in this paper is to develop a useful framework for
comparing the performance of parallel and distributed programming
systems, to help users and developers evaluate the performance tradeoffs of these systems.

Existing approaches to this problem focus on \emph{proxy-}/\emph{mini-apps}
or \emph{microbenchmarks}. These smaller codes distill key
computational characteristics of larger applications: mini-apps are
often derived from a larger code, and thus inherit some subset of its
properties, while benchmarks are typically chosen to reflect a more
narrow set of behavior(s). In either case, while a variety of insight
can be gained, the overall programming effort required is proportional to the
product of the number of systems and behaviors being
evaluated. Few published studies compare more than a
handful of systems~\cite{LULESH13, Deakin19}.

\zap{
One approach to comparing performance is through \emph{proxy-} or
\emph{mini-apps}. Because they distill key computational characteristics of larger
applications, mini-apps offer insight
without the expense of developing production codes. However, despite the name, our experience is that
mini-apps still require significant investment to develop
to the level of quality needed for useful benchmarking. In many cases,
the effort to tune for performance exceeds the effort to develop a correct implementation. As a result, implementations of mini-apps
often do not reach the level of maturity required to compare
systems. Few published studies compare more than a handful of systems~\cite{LULESH13, Deakin19}.
} 

We present Task Bench, a parameterized benchmark for exploring the performance
of parallel and distributed programming systems under a
variety of conditions.  The key property of Task Bench is that it completely separates
the system-specific implementation from the implementation
of the benchmarks themselves.
In all previous benchmarks we know of, the effort to implement $m$ benchmarks on $n$
systems is $\mathcal{O}(mn)$.  Task Bench's design reduces this work to $\mathcal{O}(m + n)$,
enabling dramatically more systems and benchmarks to be explored for the same amount of programming
effort.  New benchmarks created with Task Bench
immediately run on all systems, and new systems that implement the Task Bench interface immediately run all
benchmarks. 

Benchmarks in Task Bench are based on the observation that regardless
of the programming system in which an application is written, many
applications can be modeled as coarse-grain units of work, called
\emph{tasks}, with dependencies between tasks representing the
communication and synchronization required for parallel and
distributed execution. By explicitly modeling the \emph{task graph}
(with tasks as vertices and dependencies as edges), we make it
possible to explore a wide variety of patterns relevant to
parallel and distributed computing: trivial parallelism, halo exchanges (as
in structured and unstructured mesh codes), sweeps (as
in the discrete ordinates method of simulating radiation), FFTs, trees
(for divide and conquer), DNNs, graph analytics, etc. Tasks execute kernels with a
variety of computational properties, including compute- and
memory-bound loops of varying duration. Dependencies can be configured to carry communication payloads of varying size. Finally, multiple
(potentially heterogeneous) task graphs can be executed concurrently
to introduce task parallelism into the workload. Together, these
elements enable the exploration of a large space of application
behaviors---and make it easy to explore cases limited by runtime
overhead as well as ones where computation or communication is
dominant.

Adding a system to Task Bench involves implementing a set of standard
services, such as executing a task or data transfer. Though
benchmarks are described in terms of task graphs, this is simply a
convenient representation of the computation, and the underlying
system need not provide any native support for tasks. We provide
Task Bench implementations in systems as diverse as MPI and
Spark. Task Bench provides a core API that encapsulates functionality
shared among systems, which reduces implementation effort and makes it
much easier to achieve truly apples-to-apples comparisons between
systems.

This approach has allowed us to benchmark 15 very different parallel
and distributed programming systems (see
Table~\ref{tab:systems}).  By running all systems on common benchmarks
we were able to quantify phenomena
that have never before been measured.
Most strikingly, the overheads of systems we examine vary by more than five orders
of magnitude, with popular, widely used systems at both ends of the spectrum!  Clearly,
slower systems have ``good enough'' performance for some applications, while presumably
providing advantages in programmer productivity.

How does one predict whether performance will be good enough for a given application?
The most commonly reported measures,
\emph{weak} and \emph{strong} scaling, do not directly characterize
the performance of the underlying
programming system. Weak scaling can hide arbitrary amounts of runtime
system overhead by using sufficiently large problem sizes, and strong
scaling does not separate runtime system overhead from application costs
(such as communication) that scale with the number of nodes when
using progressively larger portions of a machine. 

To characterize the contribution of runtime overheads
to application performance, and as an example of the novel studies that can be done
with Task Bench, we introduce a new metric called
\emph{minimum effective task granularity} (METG). Intuitively, for a given
workload, METG(50\%) is the smallest task granularity that maintains
at least 50\% efficiency, meaning that the application achieves at
least 50\% of the highest performance (in FLOP/s, B/s, or other
application-specific measure) achieved on a given
machine. The efficiency bound in METG is a key innovation over
previous approaches, such as \emph{tasks per second} (TPS), that fail
to consider the amount of useful work performed (if tasks are
non-empty~\cite{Canary16, Armstrong14}) or to perform useful work at all (if tasks are empty~\cite{LegionTracing18}).

METG captures the important essence of a
weak or strong scaling study, the behavior at the limit of
scalability. For weak scaling, METG(50\%) corresponds to the
smallest problem size that can be weak-scaled with 50\%
efficiency. For strong scaling, METG(50\%) can be used to compute the
scale at which efficiency can be expected to dip below 50\%.
We note that METG(50\%) for a given runtime system will
vary with the application and the underlying hardware---i.e., METG(50\%)
is not a constant for a given system, but we find that systems have
a characteristic range of METG(50\%) values and that there is additional insight
in the reasons that METG can vary.

A lower METG does not necessarily mean that
performance for a particular workload is significantly better. Two systems with METG(50\%) of 100~\textmu{}s and 1~ms,
respectively, running an application with 10~ms average task granularity, are both likely to perform well. Only when task
granularity approaches (or drops below) METG(50\%) will they
likely diverge. METG identifies the regime in which a
given system can deliver good performance, and explains how
different systems coexist with runtime overheads that vary by orders of magnitude.

\zap{
Task Bench and METG address issues common in limit studies of runtime
systems for parallel and distributed programming. Such studies often
employ the metric \emph{tasks per second} (TPS), which is almost
universally measured with trivial (i.e., no) dependencies \cite{LegionTracing18, Canary16, Armstrong14}. While
phrased in terms of tasks, TPS can be measured for any system as long
as the application in question has identifiable units of work that run
to completion without interruption. TPS is an upper bound on
runtime-limited application throughput. But it is not a tight bound, as the
cost of nontrivial dependencies can be significant. This issue can be easily fixed by running nontrivial
configurations of Task Bench.

There is another, deeper issue with TPS. TPS may be measured with
empty tasks~\cite{LegionTracing18} or with tasks of some
duration~\cite{Canary16, Armstrong14}. When using empty tasks, the
resulting upper bound on task scheduling throughput fails to represent
useful work within a realistic application. With non-empty tasks,
\emph{where the efficiency of the overall application is not
 reported}, TPS is not a measurement of runtime-limited
performance. Large tasks may be used to hide any amount of runtime
overhead, while small tasks may result in a drop in total
application throughput even as TPS increases. Only by constraining
efficiency, as in METG, can we meaningfully measure how runtime
overhead impacts the ability to perform useful application work.
} 

\let\oldbrokenpenalty\brokenpenalty

We conduct a comprehensive study of all 15 Task Bench implementations on
up to 256 Haswell nodes of the Cori supercomputer~\cite{Cori}.
Using METG, we find that a number of factors---node
count, accelerators, and complex dependencies, among
others---individually or in combination contribute to an order of
magnitude or greater increase in METG, even in systems with the lowest
overheads. While some systems can achieve sub-microsecond METG(50\%) in
best-case scenarios, we show that a more realistic
bound for running nearly any application at scale is 100~\textmu{}s with
current technologies. Our study includes several asynchronous systems
designed to provide benefits such as overlapped computation
and communication. While small-scale benchmarks of these systems
suffer from increased overhead, we find that the benefits of these systems become
tangible at scale (provided the runtime overhead doesn't increase
beyond about 100~\textmu{}s per task).

Beyond comparative study, the ability to explore a large configuration
space also enables the discovery of bugs in the underlying systems. We
found five performance issues, ranging from communication efficiency
(Chapel, Realm), to the efficiency of task pruning, analysis and
constant folding (PaRSEC, Dask and TensorFlow). Three have been fixed
and all have been acknowledged by the developers of the respective
systems. (All were either fixed or worked around in our experiments.)
In some cases these correspond to order of magnitude or even
asymptotic improvements in the performance of the underlying
systems---benefits which apply well beyond Task Bench to all classes
of applications. The bugs are described in more detail in Section~\ref{sec:case-study}.

The paper is organized as follows: Section~\ref{sec:task-bench}
describes the Task Bench design. Section~\ref{sec:implementation}
discusses implementations in 15 systems.  Section~\ref{sec:metg}
defines METG and its relationship to quantities of interest to
application developers.  Section~\ref{sec:evaluation} provides a
comprehensive evaluation on Cori. Section~\ref{sec:case-study} describes bugs found with Task Bench. Section~\ref{sec:related-work} relates to
previous efforts; Section~\ref{sec:conclusion} concludes.
 \section{Task Bench}
\label{sec:task-bench}

\begin{table}[t]
\small

\begin{tabular}{@{} l | l | l @{}}
Parameter & Values & Purpose \\
\hline

height & height of graph & number of timesteps \\
width & width of graph & degree of parallelism \\
dependence & trivial, stencil, etc. & communication pattern \\
\quad \raisebox{0.35ex}{$\llcorner$} radix & \quad (for nearest pattern) & dependencies per task \\
kernel & compute, memory, etc. & type of kernel \\
\quad \raisebox{0.35ex}{$\llcorner$} iter. & \quad (for all kernels) & task duration \\
\quad \raisebox{0.35ex}{$\llcorner$} span & \quad (for memory kernel) & bytes used per task per iter. \\
\quad \raisebox{0.35ex}{$\llcorner$} scratch & \quad (for memory kernel) & total working set size \\
\quad \raisebox{0.35ex}{$\llcorner$} imbal. & \quad (for load imbalance) & degree of imbalance \\
output & bytes per dependency & degree of comm.
\end{tabular}

\caption{Task Bench parameters.\label{tab:parameters}}
\vspace{-0.5cm}
\end{table}
 
To explore as broad a space of application scenarios as possible, Task
Bench provides a large number of configuration parameters. These
parameters are described in
Table~\ref{tab:parameters}, and control the size and
structure of the task graph, the type and duration of kernels
associated with each task, and the amount of data associated with
each dependence edge in the graph.

\begin{figure}[t]

\subfloat[Trivial.]{
\includegraphics[width=0.28\columnwidth]{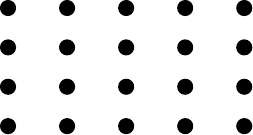}
}
\hfill
\subfloat[Stencil.\label{fig:task-graphs-stencil}]{
\includegraphics[width=0.28\columnwidth]{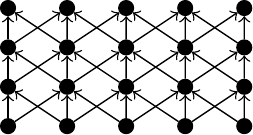}
}
\hfill
\subfloat[FFT.]{
\includegraphics[width=0.28\columnwidth]{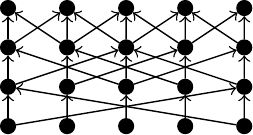}
}

{\captionsetup[subfloat]{farskip=-0.15cm,captionskip=-0.15cm}
\subfloat[Sweep.]{
\includegraphics[width=0.28\columnwidth]{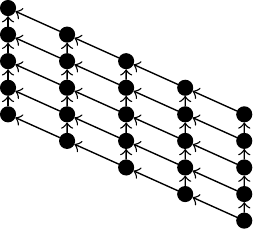}
}
\hfill
\subfloat[Tree.]{
\includegraphics[width=0.32\columnwidth]{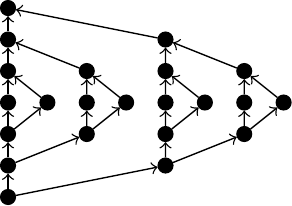}
}
\hfill
\subfloat[Random.]{
\includegraphics[width=0.28\columnwidth]{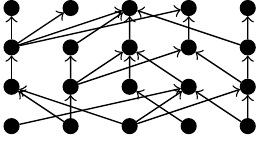}
}
}

\vspace{-0.25cm}
\caption{Sample task graphs.\label{fig:task-graphs}}
\vspace{-0.2cm}
\end{figure}

Task graphs are a combination of an \emph{iteration space} (with a task for
each point in the space) with a \emph{dependence relation}.
For simplicity, but without loss of generality, the iteration space in
Task Bench is constrained to be 2-dimensional, with time along
the vertical axis and parallel tasks along the
horizontal. Tasks may depend only on tasks from the immediately
preceding time step. Figure~\ref{fig:task-graphs} shows a number of sample task
graphs that can be implemented with Task Bench. Note that layout is
significant: generally speaking each column will be
assigned to execute on a different processor core.

Dependencies between tasks are determined by a dependence
relation. The
dependence relation identifies the tasks from the
previous time step each task depends on, permitting a wide variety
of patterns to be implemented that are relevant to real
applications: stencils,
sweeps, FFTs, trees, etc. Dependence relations may be
parameterized, such as picking the $K$ nearest neighbors, or $K$
distant neighbors. They may also vary over time, such as in the FFT pattern. The set of dependence relations is extensible, making it easy to add patterns to represent new classes of applications. Table~\ref{tab:equations} shows equations for the
dependence relations of the patterns in Figure~\ref{fig:task-graphs},
where $t$ is timestep, $i$ is column, and $W$ is the width of the task
graph.

 \begin{table}[t]
\small
\begin{tabular}{@{} l | l @{}}
Pattern & Dependence Relation \\
\hline

Trivial & $D(t, i) := \emptyset$ \\
Stencil & $D(t, i) := \{ i, i - 1, i + 1 \}$ \\
FFT & $D(t, i) := \{ i, i - 2^t, i + 2^t \}$ \\
Sweep & $D(t, i) := \{ i, i - 1 \}$ \\
Tree & $D(t, i) := \left\{ \!\!\!\! \arraycolsep=4pt\begin{array}{l l} \{ i - 2^{-t}W(i \operatorname{mod} 2^{-t + 1}W) \} & \text{if } t \le \operatorname{log}_2 W \\ \{ i, i + 2^{t-1}W^{-1} \} & \text{otherwise} \end{array} \right. \!\!\!\! $ \\
Rand. & $D(t, i) := \{ i | 0 \le i < W \wedge \operatorname{random}() < 0.5  \}$ \\
\end{tabular}

\caption{Dependence relations for sample task graphs.\label{tab:equations}}
\vspace{-0.5cm}
\end{table}
 
Listing~\ref{lst:code-sample} shows an excerpt from the Task Bench
implementation in MPI. Methods of the \lstinline{Graph} object
\lstinline{g} are provided by Task Bench's core API and are shared
among all implementations. These methods are summarized in
Table~\ref{tab:api}. The MPI implementation follows the style of
communicating sequential processes (CSP)~\cite{HoareCSP1978}, and executes a set of send
and receive calls (lines 24 and 16, respectively) followed by
executing the task body (line 32). Despite MPI having no notion of
task, the execution of a task graph maps into the CSP style in a
straightforward way. The implementation is both simple and efficient,
but due to the choice of CSP makes no attempt to exploit task
parallelism, and leaves performance on the table when executing task
graphs with load imbalance or significant communication. Note the excerpt is
simplified for presentation and the full implementation is more
general and provides additional optimizations.

In addition to specifying the shape of the task graph, the
core API also provides implementations of the kernels
executed by each task as well as other utility routines (to parse
inputs and display results). An excerpt from the core API compute kernel
is shown in Listing~\ref{lst:compute-kernel}. In addition to reducing
the effort required to implement Task Bench, providing central
implementations of these services ensures that all Task Bench
implementations can be scripted uniformly and eliminates a potential
source of performance disparity that can be a pitfall for other benchmarks.

\zap{
Despite its generality, Task Bench is easy to implement, making it
tractable to develop a suite of high-quality implementations. The central aspects of Task Bench, such as generating
task graphs and enumerating dependencies, are encapsulated in a core
library that is shared among all the Task Bench implementations. The
core library also includes implementations of the kernels, ensuring
that the kernels are identical in all systems, eliminating a potential
source of performance disparity that can be a pitfall for
implementations of mini-apps. Finally, the core library manages
parsing input parameters and displaying results,
ensuring that all implementations behave uniformly and can
be scripted consistently. Because much of the functionality needed for
a Task Bench implementation is in the core library, implementations of
Task Bench are small: our 15 Task Bench implementations range from 88
to 1500 lines, with several hundred lines being
typical. Listings~\ref{lst:compute-kernel} and \ref{lst:code-sample}
show excerpts from the Task Bench core, and the Dask implementation,
respectively. Only the code in Listing~\ref{lst:code-sample} is
implemented for each system, minimizing the work required for each additional system.
} 

The Task Bench core library is fully self-validating: The output of each
task is a tuple $\langle \textrm{row, col}\rangle$ and is unique for a given task
graph. Inputs are verified by checking the expected dependencies
against those received, and an assertion is thrown if validation
fails. These checks ensure that every execution of Task Bench is
correct. Note that the graph representation is concise, making these
checks very inexpensive. An evaluation of the performance impact of
validation showed it to be less than 3\% at the smallest task
granularities in any Task Bench implementation, with a negligible
effect on overall results.

\begin{table}[t]
\small
\begin{tabular}{@{} l | l @{}}
Method & Purpose \\
\hline
Graph::contains\_point(t,i) & is task(t,i) contained in the graph? \\
Graph::deps(t,i) & predecessors of task(t,i) \\
Graph::reverse\_deps(t,i) & successors of task(t,i)\\
Graph::execute\_point(t,i, ...) & execute the body of task(t,i) \\
\end{tabular}

\caption{Subset of the core API used in code samples below.\label{tab:api}}
\vspace{-0.25cm}
\end{table}
 \begin{lstlisting}[language=C,caption={Core API implementation of compute kernel.},label={lst:compute-kernel},style=codeblock,float]
void compute_kernel(long iterations) {
  double A[64];
  for (int i = 0; i < 64; i++) A[i] = 1.2345;
  for (long iter = 0; iter < iterations; iter++)
    for (int i = 0; i < 64; i++)
      A[i] = A[i] * A[i] + A[i];
}
\end{lstlisting}
 \begin{lstlisting}[language=C++,caption={Excerpt from Task Bench implementation in MPI.},label={lst:code-sample},style=codeblocktop,float]
void execute_task_graph(Graph g) {
  char *output = (char *)malloc(g.output_bytes);
  char *scratch = (char *)malloc(g.scratch_bytes);
  char **inputs = (char **)malloc(/*...*/);
  long rank;
  // initialize data structures...

  std::vector<MPI_Request> requests;
  for (long t = 0; t < g.height; ++t) {
    if (g.contains_point(t, rank)) {
      long idx = 0;
      requests.clear();

      for (long dep : g.deps(t, rank)) {
        MPI_Request req;
        MPI_Irecv(inputs[idx], g.output_bytes, MPI_BYTE,
          dep, 0, MPI_COMM_WORLD, &req);
        requests.push_back(req);
        idx++;
      }

      for (long dep : g.reverse_deps(t, rank)) {
        MPI_Request req;
        MPI_Isend(output, g.output_bytes, MPI_BYTE,
          dep, 0, MPI_COMM_WORLD, &req);
        requests.push_back(req);
      }

      MPI_Waitall(requests.size(), requests.data(),
        MPI_STATUSES_IGNORE);

      g.execute_point(t, rank, output, inputs, scratch);
    }
  }
}
\end{lstlisting}
 
Task Bench provides two main kernels that can be called from tasks:
compute- and memory-bound. The compute-bound
kernel executes a tight loop and is hand-written using AVX2 FMA
intrinsics. The memory-bound kernel performs sequential reads and
writes over an array, again with AVX2
intrinsics. The duration of both kernels can be configured by setting
the number of iterations to execute; we use this ability to simulate
the effects of varying application problem sizes. The memory-bound
kernel is carefully written to keep the working set size constant as
the number of iterations decreases, to avoid unwanted speedups due to
cache effects.
 \section{Implementations}
\label{sec:implementation}

We have implemented Task Bench in the 15 parallel and distributed
programming systems listed in Table~\ref{tab:systems}. These include
traditional HPC programming models (MPI and MPI+X), PGAS and actor
models (Chapel, Charm++ and X10), task-based systems (OmpSs, OpenMP
4.0, PaRSEC, Realm, Regent, and StarPU) and systems for large scale
data analytics, machine learning and workflows (Dask, Spark, Swift/T,
and TensorFlow). Implementing Task Bench in such a wide range of
systems is possible because the separation between core API (in
Table~\ref{tab:api}) and system implementation enables an overall
effort of $\mathcal{O}(m + n)$ (for $m$ benchmarks on $n$ systems)
rather than $\mathcal{O}(mn)$ as has been the case for all previous
benchmarks that we know of. We briefly describe the systems and
implementations below.

One challenge in targeting such a wide variety of
systems is that the capabilities of the systems vary considerably. For
example, some systems are \emph{implicitly parallel}, and provide some
form of parallelism discovery from sequential programs, whereas others
are \emph{explicitly parallel} and require users to specify the
parallelism in the program. For systems that provide both implicit and explicit parallelism, the form of parallelism used in Task Bench is emphasized in Table~\ref{tab:systems}.

In all cases, members of the programming systems' teams
were consulted in the development and evaluation of the
corresponding Task Bench implementations. Where assistance was provided, the insights helped ensure that we provide the highest quality implementations for each system.

\subsection{Traditional HPC Programming Models}

MPI~\cite{MPI} is a message-passing API for HPC. We provide an MPI
implementation written in the style of communicating sequential
processes (CSP). An excerpt is shown in Listing~\ref{lst:code-sample}.

We provide two MPI+X implementations to evaluate hierarchical
programming models. Our MPI+OpenMP implementation uses forall-style
parallel loops to execute tasks, but otherwise follows the CSP
implementation above. The code uses shared memory for data movement
within a rank. Our MPI+CUDA implementation follows an offload model
where data is copied to and from the GPU on every timestep.

\subsection{PGAS and Actor Models}

PGAS and actor models, such as Chapel~\cite{Chapel15},
Charm++~\cite{Charmpp93} and X10~\cite{X1005} offer
asynchronous tasks, making them amenable to a straightforward
implementation of task bench. Synchronization is explicit and may be
provided by messages (in actor models) or other primitives, such as
locks or atomics (in PGAS models). PGAS models such as Chapel and X10
provide global references to data anywhere in the machine, but vary in
whether data can be accessed remotely or not (Chapel allows this, X10
does not). Chapel also provides support for implicit parallelism which
we do not evaluate in this paper.

\subsection{Task-Based Programming Models}

Task-based systems include OmpSs~\cite{OmpSs11}, OpenMP
4.0~\cite{OpenMPSpec40}, PaRSEC~\cite{PARSEC13, PARSEC_DTD},
Realm~\cite{Realm14}, Regent~\cite{Regent15}, and
StarPU~\cite{StarPU11}. Though the details vary, these systems
typically provide implicit parallelism, where tasks are enumerated
sequentially (in program order) and the dependencies between tasks are
analyzed automatically to construct a \emph{dependence graph} that
guides the execution of tasks. PaRSEC provides two modes,
\emph{dynamic task discovery} (DTD)~\cite{PARSEC_DTD}, which operates
as above, and \emph{parameterized task graphs} (PTG)~\cite{PARSEC13},
where the dependence graph is constructed automatically from an
analytical representation of the task graph. Realm (unlike the others
above) is explicitly parallel and requires tasks to be connected
explicitly via \emph{events}. Realm is the low-level execution engine for Regent and thus serves as a limit study of what can be achieved with Regent. StarPU, in addition to its usual,
implicitly parallel mode, also provides a mode where MPI is used for
synchronization (and is thus explicitly parallel).

Among implicitly parallel, distributed task-based systems, there can be a scalability
bottleneck due to enumerating tasks sequentially. PaRSEC and StarPU
allow users to manually prune the task graph, skipping tasks not
mapped for execution onto a given node, plus a ``halo'' consisting of
tasks connected via dependencies to the set of node-local tasks. Because the dynamic
checks to see if a task should be executed are not free of cost, we
also provide versions of the PaRSEC and StarPU implementations
(labeled \lstinline{shard} and \lstinline{expl}, respectively)
hand-written to minimize such costs. PaRSEC \lstinline{shard} uses the
DTD mode but manually minimizes dynamic checks. StarPU
\lstinline{expl} uses the MPI integration described above. We see in
Section~\ref{subsec:scalability} that such modifications are needed to
achieve optimal scalability. Regent performs an equivalent
optimization at compile time~\cite{ControlReplication17} that does not
require user intervention and preserves the original, implicitly
parallel programming model of the language.

\begin{table}[t]
\small
\begin{tabular}{@{} l | l | l | l | l @{}}
System & Paradigm & Parallelism & Distrib. & Network \\
\hline
Chapel & multi-resolution & \emph{expl.}, impl. & yes & uGNI\footnote{Chapel uses GASNet to support non-Cray networks.} \\
Charm++ & actor model & explicit & yes & uGNI\footnote{Charm++ provides additional backends for other networks.} \\
Dask & task-based & implicit & yes & sockets \\
MPI & message passing & explicit & yes & uGNI\footnote{Most MPI implementations provide additional backends for other networks.} \\
MPI+X & hybrid & explicit & yes & MPI \\
OmpSs & loop-, task-based & expl., \emph{impl.} & no & \\
OpenMP & loop-, task-based & expl., \emph{impl.} & no & \\
PaRSEC & task-based & implicit & yes & MPI \\
Realm & task-based & explicit & yes & GASNet \\
Regent & task-based & implicit & yes & GASNet \\
Spark & functional & implicit & yes & sockets \\
StarPU & task-based & \emph{expl., impl.} & yes & MPI \\
Swift/T & dataflow & implicit & yes & MPI \\
TensorFlow & dataflow & explicit & yes\footnote{Our evaluation only considers TensorFlow on a single node.} & sockets \\
X10 & place-based & explicit & yes & MPI\footnote{X10 also provides a PAMI backend on supported networks.}
\end{tabular}

\caption{Systems for which we implemented Task Bench.\label{tab:systems}}
\vspace{-0.5cm}
\end{table}
 
\subsection{Data Analytics, Machine Learning and Workflows}

Dask~\cite{Dask15}, Spark~\cite{Spark10} and
TensorFlow~\cite{TensorFlow15} are programming models for large scale
data analytics and machine learning. Dask and TensorFlow provide
domain-specific abstractions built on top of task-based runtimes. Our
implementations directly create tasks and are similar to other
task-based systems above.

Spark provides support for functional operators that implicitly map to
tasks. We use \lstinline[language=Scala]{flatMap} and
\lstinline[language=Scala]{groupByKey} to generate dependencies and
\lstinline[language=Scala]{mapPartitions} to execute tasks. An
explicit hash partitioner ensures the correct task granularity.

Swift/T~\cite{Wozniak13} is a parallel scripting language
with dataflow semantics, used primarily for workflow automation. Our
implementation is straightforward.
 \section{METG}
\label{sec:metg}

\begin{figure}[t]
\centering
\includegraphics[width=\columnwidth]{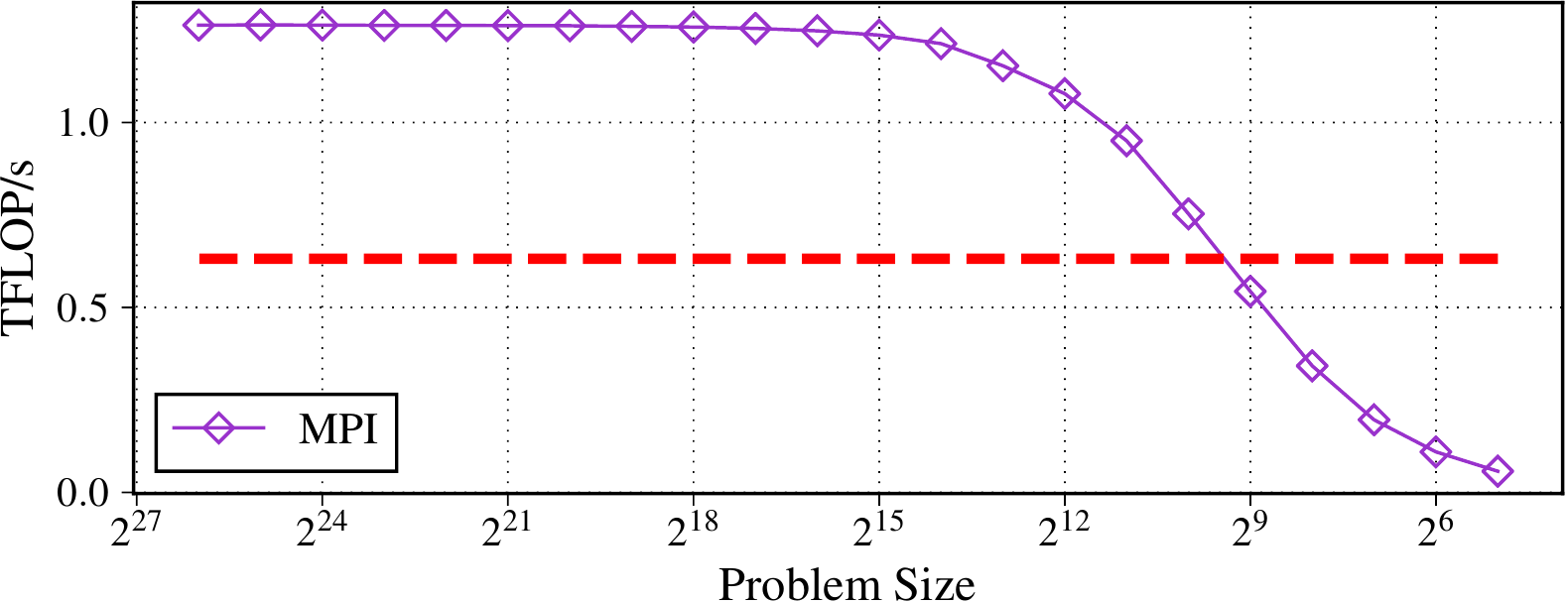}
\vspace{-0.6cm}
\caption{MPI FLOP/s vs problem size (stencil, 1 node).\label{fig:flops-mpi}}
\vspace{-0.1cm}
\end{figure}
 \begin{figure}[t]
\centering
\includegraphics[width=\columnwidth]{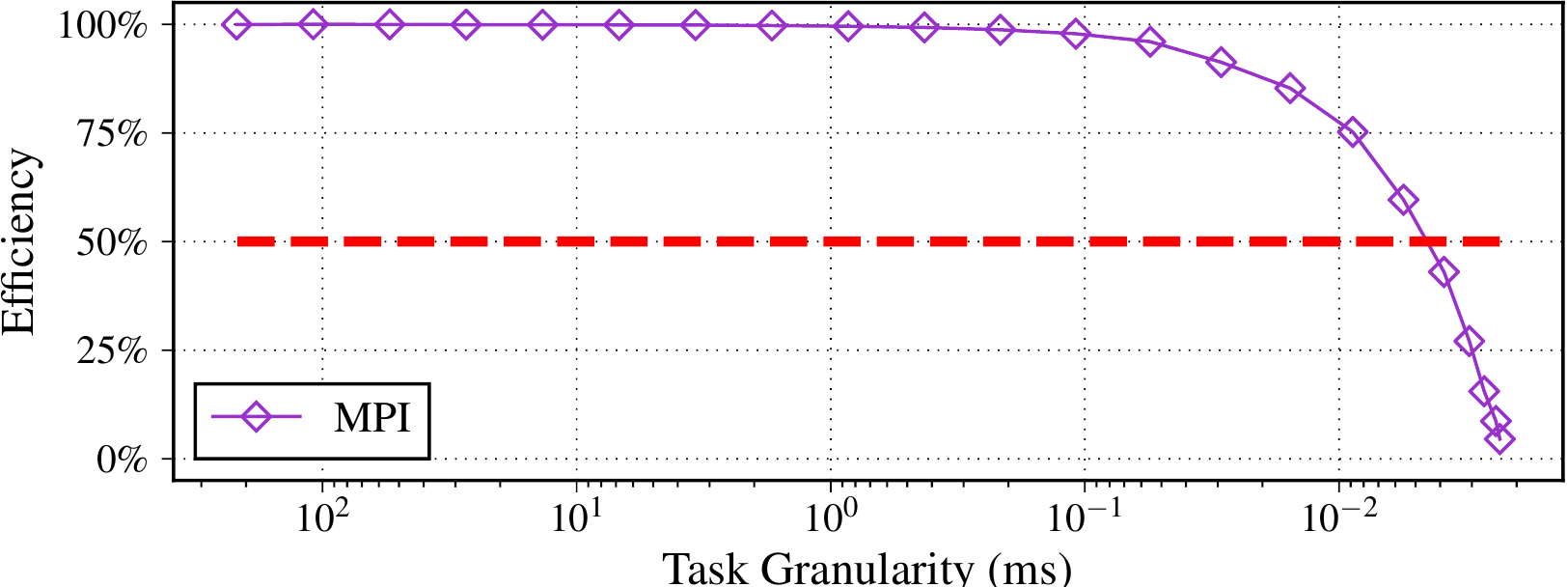}
\vspace{-0.6cm}
\caption{MPI efficiency vs task granularity (stencil, 1 node).\label{fig:efficiency-mpi}}
\vspace{-0.15cm}
\end{figure}
 
Since Task Bench permits rapid exploration of a large space
of application scenarios, one question is how to characterize the
performance and efficiency of systems under study. As noted
above, the overheads of the systems we consider vary by more than
five orders of magnitude, making it challenging to extract useful
information from weak and strong scaling runs.

Existing studies of system efficiency
typically report \emph{tasks per second} (TPS). TPS results
are difficult to interpret and apply, because efficiency (and thus
the amount of useful work) is not constrained. With empty
tasks~\cite{LegionTracing18}, the resulting upper bound on task
scheduling throughput fails to represent useful work within a
realistic application. With non-empty tasks, since the efficiency of
the overall application is typically not reported~\cite{Canary16,
  Armstrong14}, TPS is not a measurement of runtime-limited
performance. Large tasks may be used to hide any amount of runtime
overhead, while small tasks may result in a drop in total application
throughput even as TPS increases.

We introduce \emph{minimum effective task granularity}, or METG, an
efficiency-constrained metric for runtime-limited
performance. METG(50\%) for an application $A$ is
the smallest average task granularity (i.e., task duration) such that $A$
achieves overall efficiency of at least 50\%. Note that METG is parameterized by the efficiency metric. For example, in
compute-bound applications efficiency can be measured as the
percentage of the available FLOP/s achieved. On Cori with 1.26 TFLOP/s available per Haswell node, METG(50\%) corresponds to
the smallest task granularity achieved while maintaining at least 0.63
TFLOP/s per node. However, METG is not tied to peak performance, and
in applications not amenable to being characterized in this way,
another application-specific measure of performance can be used. For
example, a simulation on a mesh might use the number of mesh cells
processed per second (i.e., total number of cells divided by wall
clock time per iteration of the main simulation loop).

The choice of 50\% is a parameter and not fundamental to METG. We use
50\% in our studies to avoid pathologies associated with lower
thresholds (see Section~\ref{subsec:peak-performance-and-efficiency}),
and also because it aligns with what we observe in practice. For
example, one supercomputer center instructs users applying for
projects to run a strong scaling study and then ``select the most
parallel efficient job size,'' i.e., the largest number of nodes with
a ``ratio of benchmark speed-up vs. linear speed-up above 50\%''
\cite{CSCSReport}, which corresponds to METG(50\%).

\begin{figure}[t]
\centering
\includegraphics[width=\columnwidth]{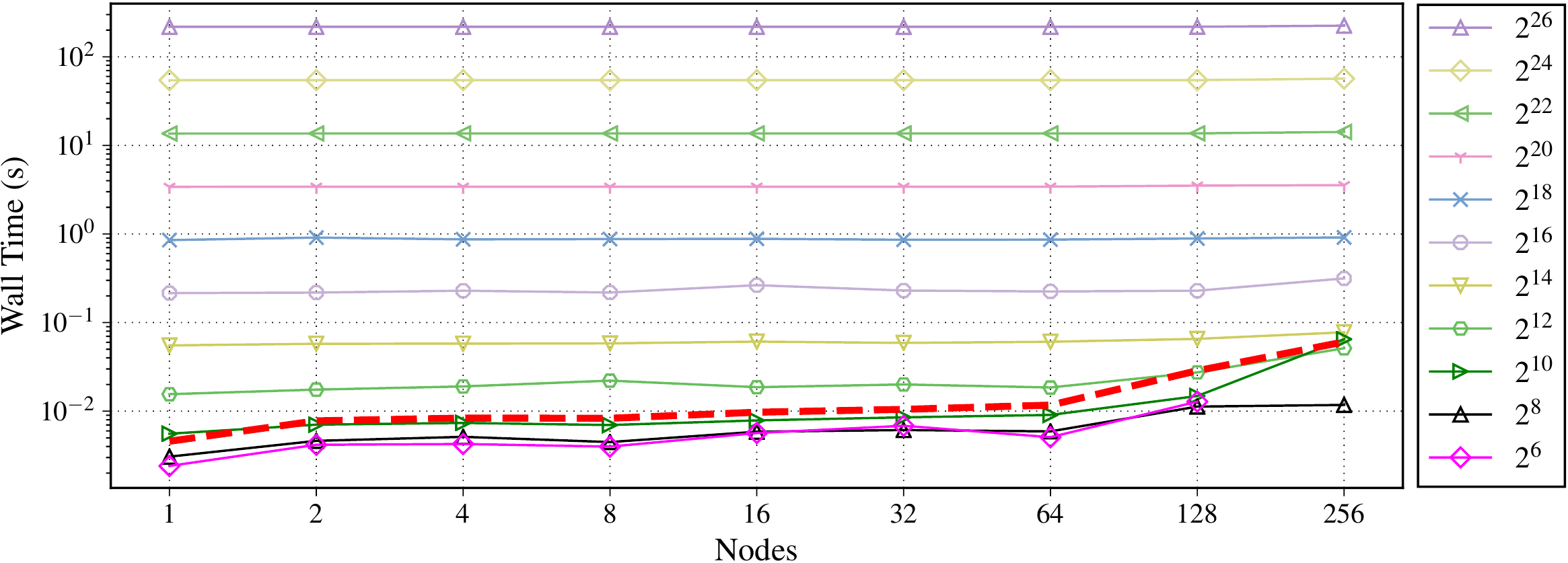}
\vspace{-0.6cm}
\caption{MPI weak scaling with problem size per node (stencil).\label{fig:weak-scaling-mpi}}
\vspace{-0.1cm}
\end{figure}
 \begin{figure}[t]
\centering
\includegraphics[width=\columnwidth]{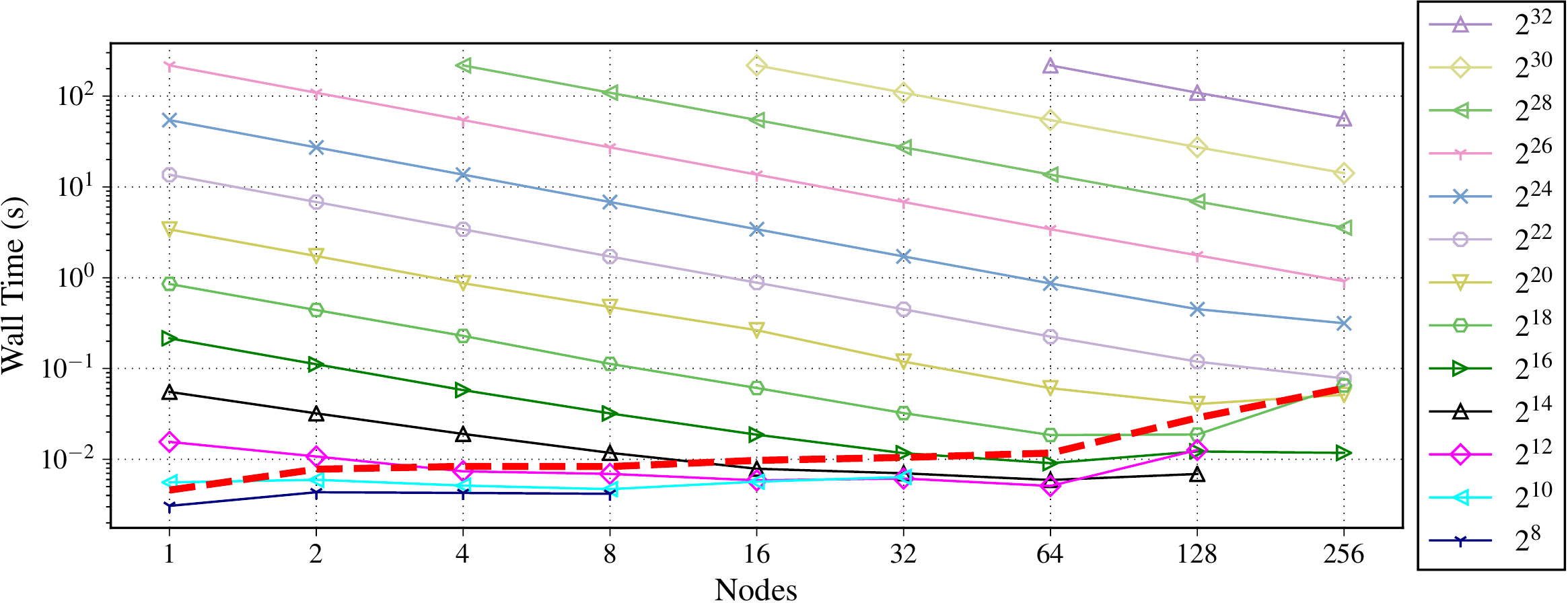}
\vspace{-0.6cm}
\caption{MPI strong scaling with problem size (stencil).\label{fig:strong-scaling-mpi}}
\vspace{-0.1cm}
\end{figure}
 
Figure~\ref{fig:flops-mpi} shows how METG is
measured. We run the application (MPI Task Bench) on a Cori Haswell node
with a problem size large enough that runtime is dominated by kernel
execution. This result confirms
that the application is properly configured and that the
efficiency metric is achievable. The problem
size is then repeatedly reduced while maintaining exactly the same hardware and software
configuration (in particular, the same number of nodes and tasks). The
expectation is that as problem size shrinks,
performance will begin to drop and eventually approach zero. Systems
with lower runtime overheads maintain higher performance at smaller
problem sizes compared to systems with higher overheads.

To calculate METG, the data is replotted along axes of efficiency
(i.e., as a percentage of the peak FLOP/s achieved) and task
granularity (i.e., $\text{wall time} \times \text{num.~cores}/\text{num.~tasks}$), as shown in Figure~\ref{fig:efficiency-mpi}. Note that a \emph{task} is defined
broadly to be any continuously-executing unit of application code,
and thus it makes sense to discuss tasks even in systems
with no explicit notion of tasking, such as MPI. In this case, the
tasks run the compute-bound kernel shown in
Section~\ref{sec:task-bench}.

In Figure~\ref{fig:efficiency-mpi},
efficiency starts at 100\%. Initially task granularity
shrinks with minimal change in efficiency. As tasks shrink further, efficiency drops more rapidly, approaching a vertical asymptote as overhead comes to dominate useful work.

METG(50\%) is the intersection of the curve at 50\% efficiency, as
shown by the red, dashed lines in Figures~\ref{fig:flops-mpi} and
\ref{fig:efficiency-mpi}. At 50\% efficiency, MPI achieves an average
task granularity of
4.6 \textmu{}s, thus the METG(50\%) of MPI is 4.6 \textmu{}s in this
configuration. 

METG has a well-defined
relationship with quantities of interest such as
weak and strong scaling. Figures~\ref{fig:weak-scaling-mpi} and
\ref{fig:strong-scaling-mpi} show the weak and strong scaling of MPI Task Bench running a stencil pattern at a variety of problem sizes. In these
figures, the vertical axis is shown as wall time to emphasize the
relationship to time-to-solution, but it could equivalently be shown
as task granularity (as the number of tasks per execution is
fixed). Intuitively, at
larger problem sizes MPI is perfectly efficient. This can be seen at
the top of each figure, with flat lines when weak scaling and
ideally-sloped downward lines when strong scaling. Inefficiency begins
to appear at smaller problem sizes, towards the bottom of the graph,
where lines become more compressed. At the
very bottom, the lines compress together as running time becomes
dominated by overhead. Note that the contour of the bottom of each
graph is identical and conforms to the METG curve (marked by the red,
dashed line).

\begin{table}[t]
\small
\begin{tabular}{@{} l | l | l @{}}
System & Version & Notes \\
\hline
Chapel & 1.18.0 & {\lstinline!--fast!} \\
Charm++ & 6.9.0 & {\lstinline!-optimize!} \\
Dask & 1.1.5 & \\
MPI(+X) & Cray MPICH 7.7.3 & {\lstinline!-O3!} \\
OmpSs & 2, release 2020.06 & {\lstinline!-O3!} \\
OpenMP & Intel KMP 18.0.1.163 & {\lstinline!-O3!} \\
PaRSEC & Git {\lstinline!master!} (\lstinline!242498d!) & {\lstinline!-O3!} \\
Realm & Git {\lstinline!subgraph!} (\lstinline!5e9dcfa!) & {\lstinline!-O3!} \\
Regent & Git {\lstinline!subgraph!} (\lstinline!5e9dcfa!) & {\lstinline!-fflow-spmd 1!} \\
Spark & 2.3.0 (Scala 2.11.8, Java 8) & \\
StarPU & 1.3.4 & {\lstinline!-O3!} \\
Swift/T & 1.4 & {\lstinline!-O3!} \\
TensorFlow & 2.1.0 & \\
X10 & Git {\lstinline!master!} (\lstinline!9212dc2!) & {\lstinline!-O3 -NO_CHECKS!}
\end{tabular}

\caption{System version and configuration notes.\label{tab:flags}}
\vspace{-0.5cm}
\end{table}
 
METG therefore has a direct relationship with the smallest problem
size that can be weak scaled to a given node count with a given level
of efficiency. Using the formula for task granularity above, each run
is 32 tasks wide and 1000 timesteps long, so task granularity is wall
time divided by 1000 (since Cori has 32 cores per node). The $2^{12}$
problem size in Figure~\ref{fig:weak-scaling-mpi} scales well
initially because the task granularity of 20~\textmu{}s is greater
than the METG(50\%) of MPI at small node counts (which is about 4.6-12
\textmu{}s from 1-64 nodes) but not at higher node counts (which rises
to 28 \textmu{}s at 128 nodes and 61 \textmu{}s at 256). Similarly,
METG corresponds to the point at which strong scaling can be expected
to stop. In Figure~\ref{fig:strong-scaling-mpi} the problem size
$2^{18}$ strong scales to 64 nodes, the point at which the
scaling curve intersects METG(50\%).

The METG metric has another useful property. Because METG is measured ``in place'' (i.e.,
without changing the number of nodes or cores available to the
application), METG isolates effects
due to shrinking problem size from effects due to
increased communication and other resource issues as
progressively larger portions of the machine are used.
 \section{Evaluation}
\label{sec:evaluation}

We present a comprehensive evaluation of our Task Bench implementations on up to 256
Haswell nodes of the Cori supercomputer~\cite{Cori}, a Cray XC40
machine. Cori Haswell nodes have 2 sockets with Intel Xeon E5-2698 v3
processors (a total of 32 physical cores per node), 128 GB RAM, and a
Cray Aries interconnect. We use GCC 7.3.0 for all Task Bench
implementations, and (where applicable) the system default MPI
implementation, Cray MPICH 7.7.3. Versions and flags for the
various systems are shown in Table~\ref{tab:flags}.

\begin{figure}[t]
\centering
\includegraphics[width=\columnwidth]{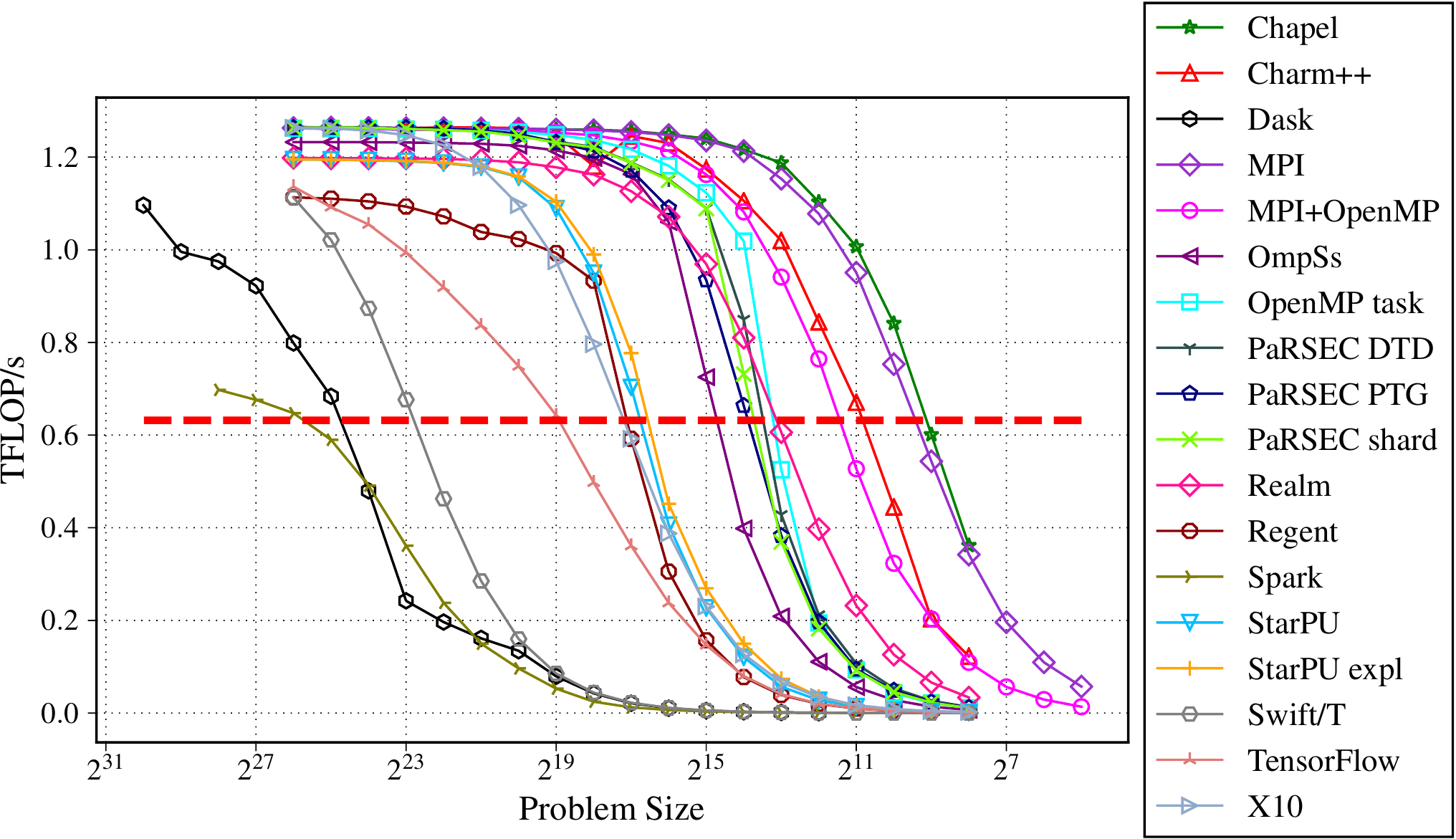}
\vspace{-0.6cm}
\caption{FLOPS vs problem size (stencil, 1 node). Higher is better.\label{fig:flops}}
\vspace{-0.35cm}
\end{figure}
 \begin{figure}[t]
\centering
\includegraphics[width=\columnwidth]{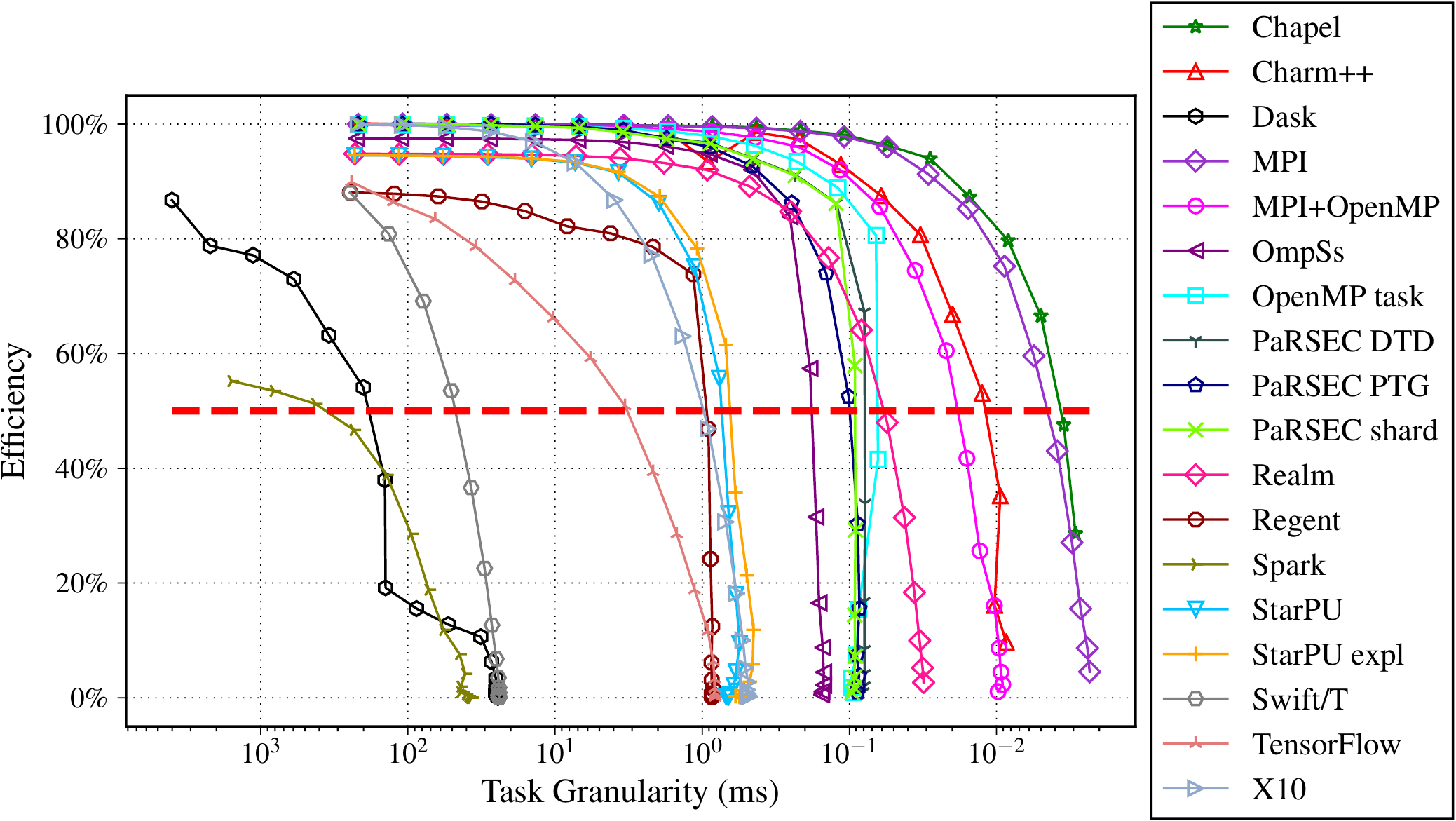}
\vspace{-0.6cm}
\caption{Efficiency vs task granularity (stencil, 1 node). Higher is better.\label{fig:efficiency}}
\vspace{-0.35cm}
\end{figure}
 \begin{figure}[t!]
\centering
\includegraphics[width=\columnwidth]{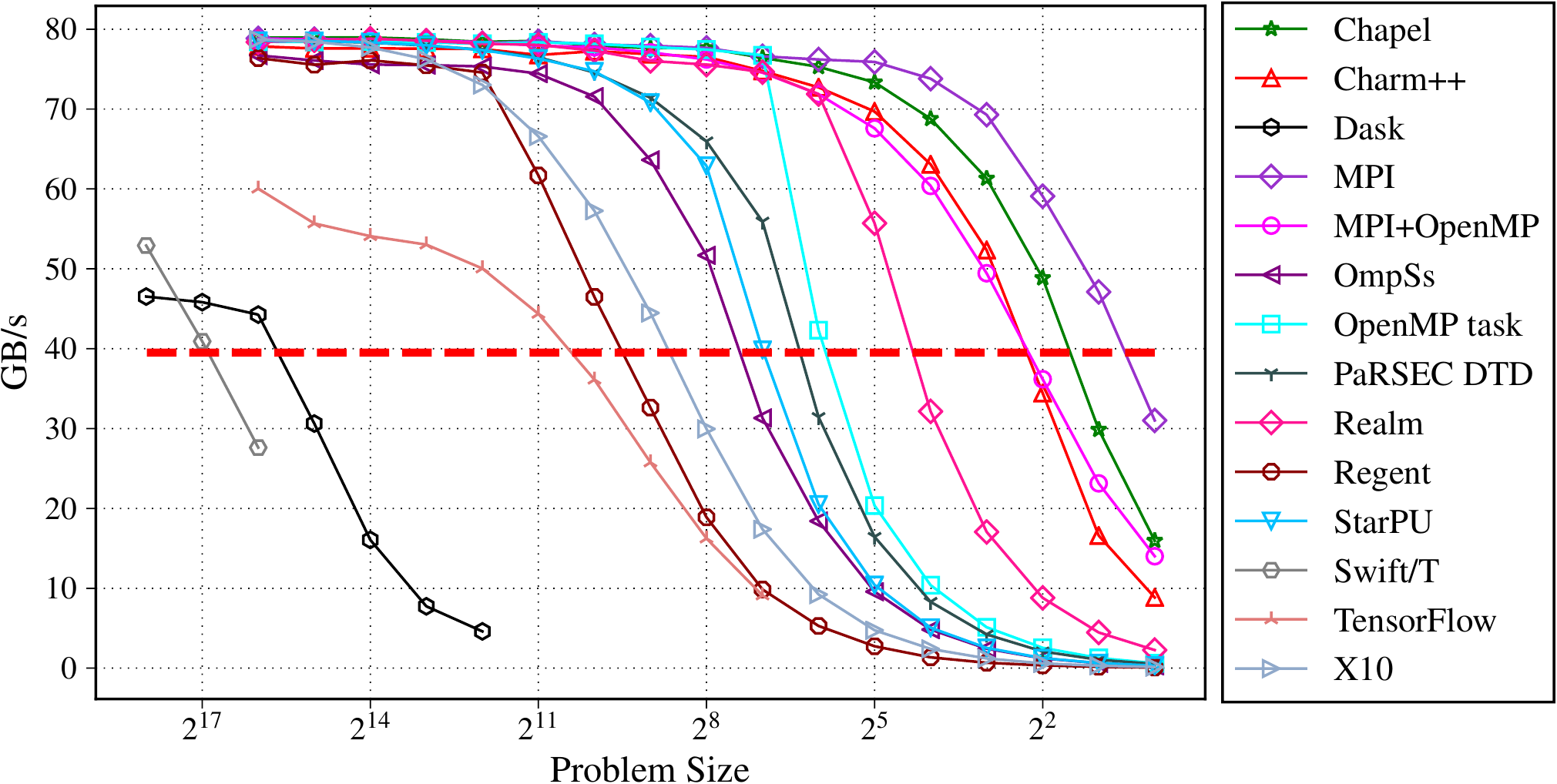}
\vspace{-0.6cm}
\caption{B/s vs problem size (stencil, 1 node). Higher is better.\label{fig:bytes}}
\vspace{-0.35cm}
\end{figure}
 
\begin{figure*}[t]

{\captionsetup[subfloat]{farskip=-0.1cm}
\subfloat[Stencil pattern.\label{fig:metg-compute-stencil}]{
\includegraphics[width=\columnwidth]{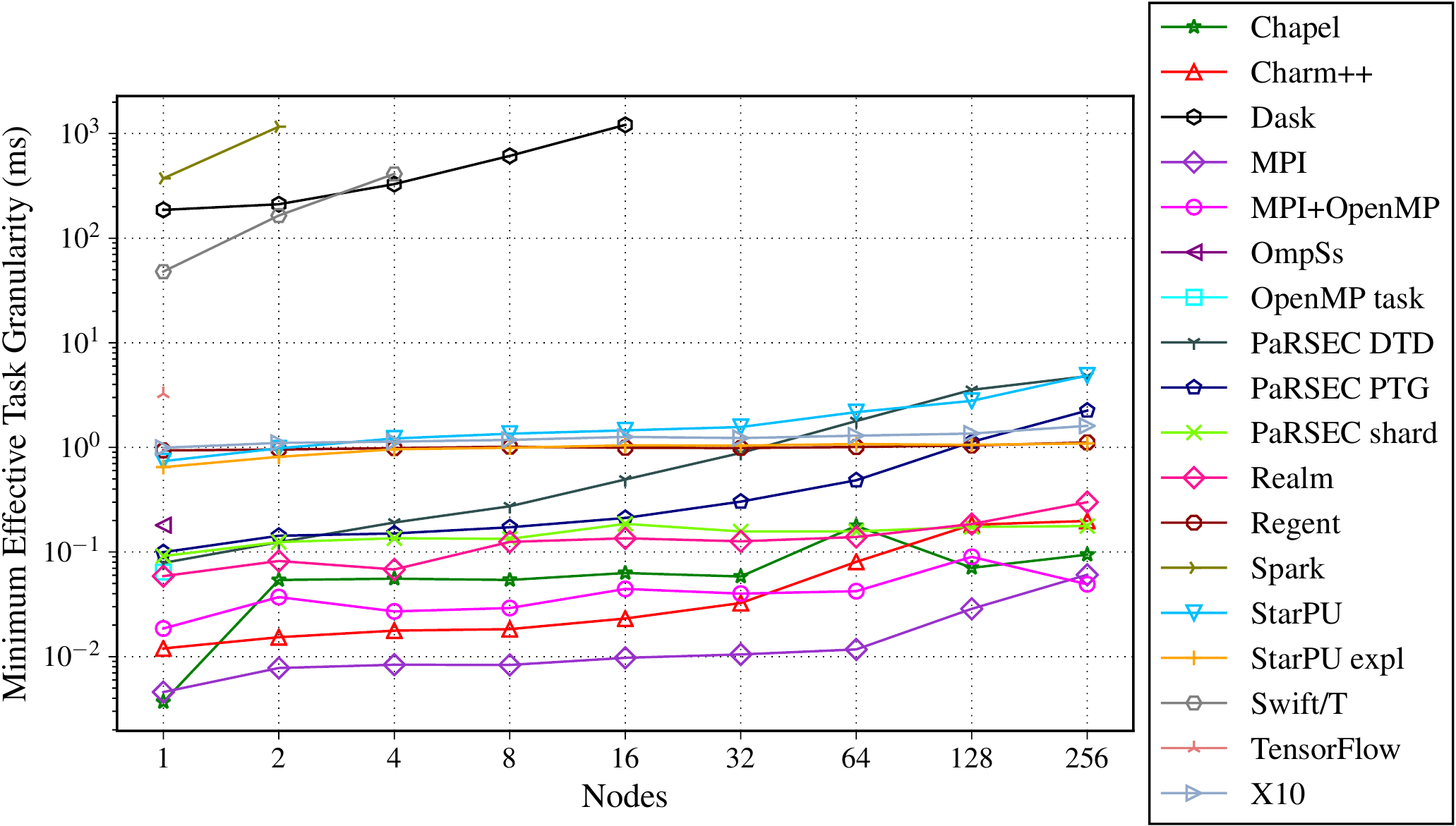}
}
\subfloat[Nearest pattern, 5 deps/task.\label{fig:metg-compute-nearest}]{
\includegraphics[width=\columnwidth]{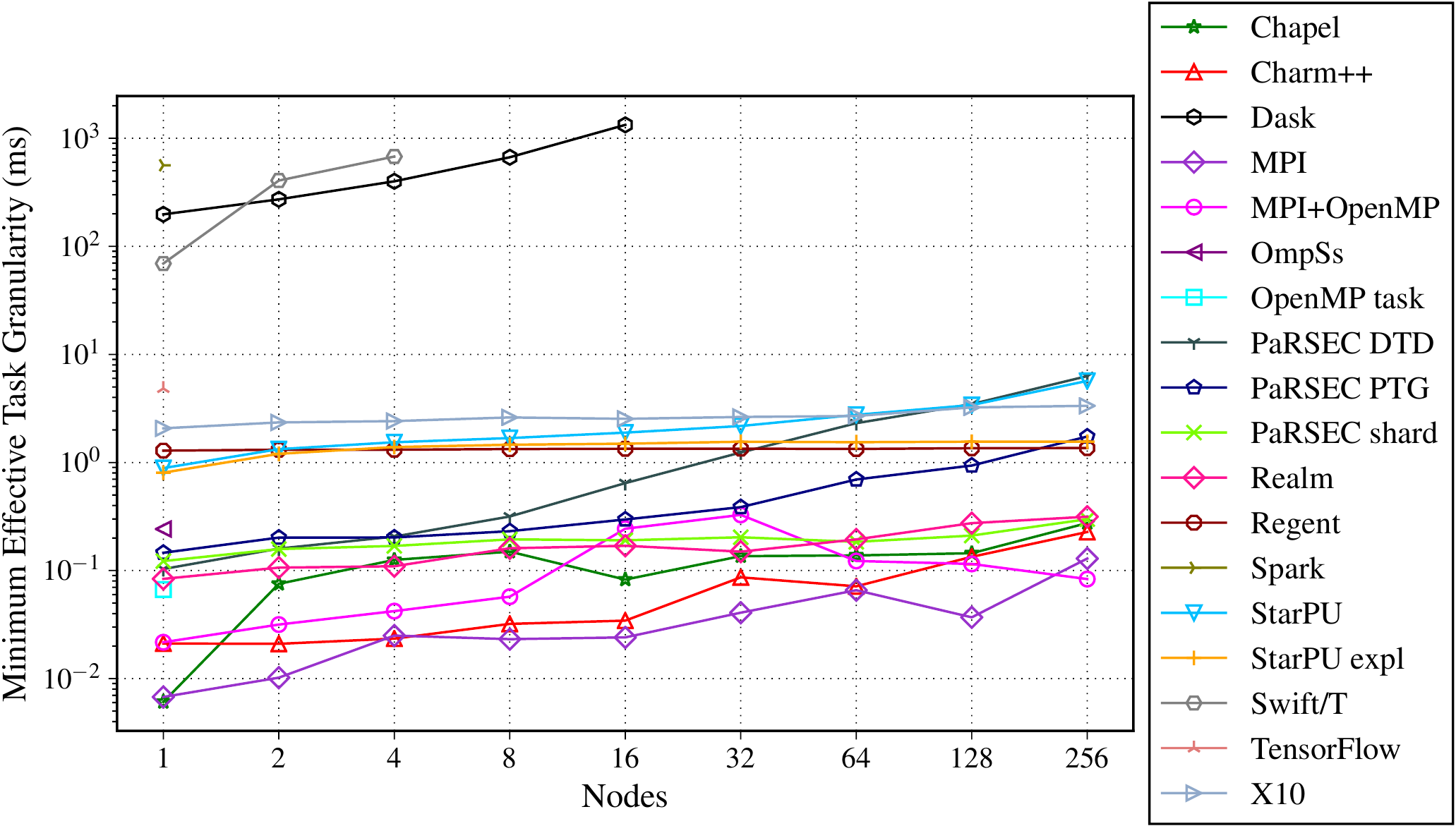}
}
}

\subfloat[Spread pattern, 5 deps/task.\label{fig:metg-compute-spread}]{
\includegraphics[width=\columnwidth]{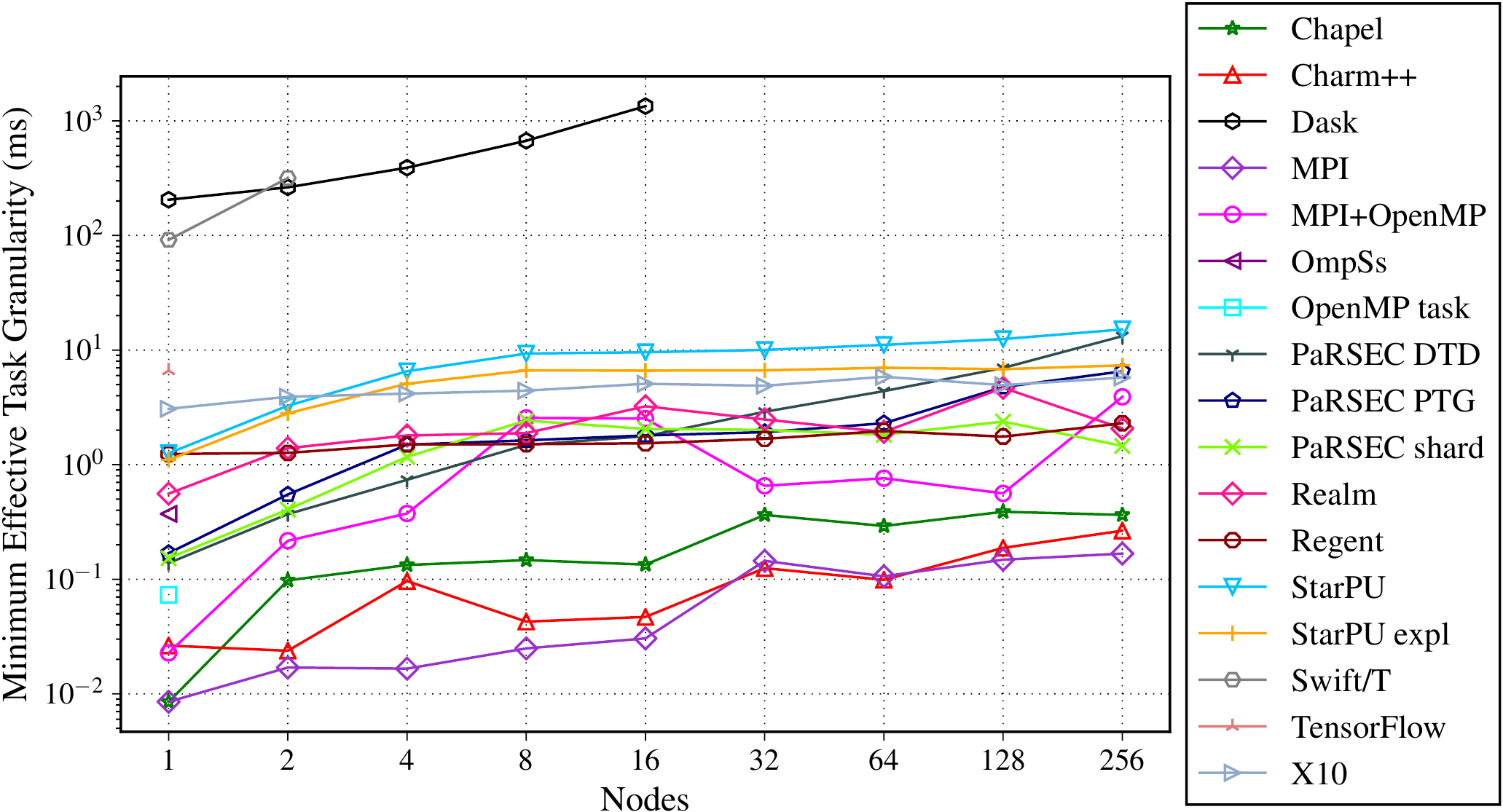}
}
\subfloat[Nearest pattern, 5 deps/task, 4 independent graphs.\label{fig:metg-compute-4x-nearest}]{
\includegraphics[width=\columnwidth]{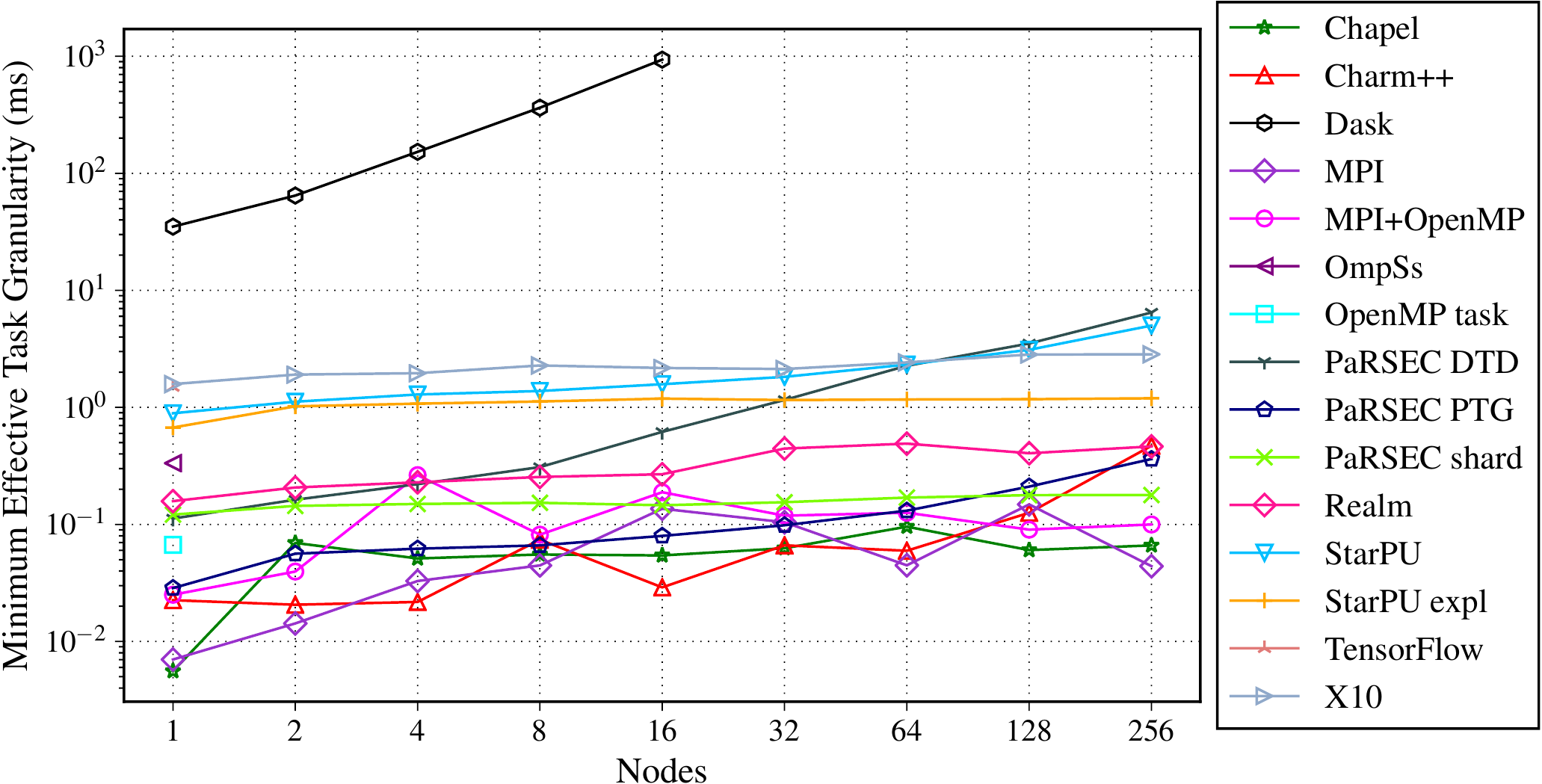}
}

\vspace{-0.15cm}
\caption{METG vs node count for different dependence patterns. Lower is better.\label{fig:metg-compute}}
\vspace{-0.25cm}
\end{figure*}
 
For GPU experiments we use Piz Daint~\cite{PizDaint}, a Cray XC50 with
a Intel Xeon E5-2690 v3 (12 physical cores) and one NVIDIA Tesla
P100 per node. We use GCC 6.2.0, Cray MPICH 7.7.2, and
CUDA 9.1.85.

\subsection{Compute Kernel Performance}
\label{subsec:peak-performance-and-efficiency}

We first consider the peak performance achieved by each system. There
should exist some task granularity which is sufficient to offset the
runtime overheads of any system, regardless of how large those
are. Even so, a variety of issues can lead to performance loss
(e.g. due to not using all available cores). Verifying that peak
performance is achieved ensures that there are no such flaws in our
configuration.

Figure~\ref{fig:flops}, which is the full version of
Figure~\ref{fig:flops-mpi}, shows the FLOP/s achieved with a compute-bound
kernel with varying problem sizes (simulated by running the kernel for varying numbers of iterations). Each data point in the graph is a mean of 5 runs, with Task Bench configured to execute 1000 time steps of the stencil pattern. In the best case, we measure peak FLOP/s of
$1.26 \times 10^{12}$, which compares favorably with the officially
reported number of $1.2 \times 10^{12}$ \cite{Cori}. We use our empirically determined number as
the baseline for 100\% efficiency below.

Most systems achieve or nearly achieve peak FLOP/s. Some
systems reserve a number of cores (usually 1 or 2) for internal
use (see below); these systems take a minor hit in peak FLOP/s compared
to systems that share all cores between application and runtime. Some
higher-overhead systems struggle to achieve peak FLOP/s, though in most cases
the curves suggest that performance would continue to improve if we
were to run larger problem sizes. Unfortunately, the excessive
computational cost of running such tests makes this prohibitively
expensive. For example, the Spark job in
this case ran for over 6 hours.

Figure~\ref{fig:efficiency} plots efficiency (as a percentage of
peak FLOP/s) vs. task granularity. As described in Section~\ref{sec:metg}, this is
used to calculate METG(50\%). The
red, dashed line indicates 50\% efficiency.
In most cases, task granularity asymptotes prior to this point,
though some systems continue to improve at lower values. Accounting for this effect is one of the main arguments in
favor of using reasonable efficiency thresholds for METG instead of
empty tasks
(i.e., METG(0\%)). Empty tasks
reward strategies, such as devoting 100\% of
system resources to the runtime system, that make no sense for real
applications.

\subsection{Memory Kernel Performance}

Figure~\ref{fig:bytes} shows performance with a memory-bound kernel. We measure a peak memory
bandwidth of 79~GB/s, using a working set size of 0.5 GB. As discussed in Section~\ref{sec:task-bench}, the kernels are designed to keep the working set constant as the number of iterations decrease to avoid noisy, superlinear effects in the results. For comparison, the
OpenMP-enabled STREAM benchmarks~\cite{STREAM} report up to 98~GB/s on the same hardware.

Not all cores are required to saturate memory bandwidth.
  This reduces the impact of reserving cores for system use
  (e.g. task-based systems that perform dependence analysis). Nearly
all systems hit 100\% of peak, unlike the compute-bound case.

The remaining experiments use compute-bound kernels.

\subsection{Baseline Overhead}

One question when considering different programming
systems is: How much overhead does the system add? This question is tricky to answer directly because some systems introduce
overhead \emph{inline} (i.e., by running system internal processes on
the same cores as application tasks), while other systems introduce
overhead \emph{out-of-line} (i.e., by dedicating one or more cores
solely to runtime use). Some systems, like Charm++, PaRSEC, Realm, and Regent,
support both configurations.

To answer this question, we use METG as a proxy for overhead. Figure~\ref{fig:metg-compute}
shows how METG(50\%) varies with node count for a subset of
dependence patterns supported by Task Bench. METG(50\%) is calculated
separately at each node count, to distinguish runtime system behavior from
changes in communication latency and topology when using
progressively larger portions of the machine.

We consider the following configurations of Task Bench:
Figure~\ref{fig:metg-compute-stencil} is a 1D stencil where each task
depends on 3 other tasks (including the same point in the previous
timestep). Figure~\ref{fig:metg-compute-nearest} is a pattern where
each task depends on 5 others, chosen to be as close as
possible. Figure~\ref{fig:metg-compute-spread} is a pattern where each
task depends on 5 others, spread as widely as possible. And
Figure~\ref{fig:metg-compute-4x-nearest} shows 4 identical copies of
the nearest dependence pattern executing concurrently.

We observe that overheads vary by over 5 orders of
magnitude. The most efficient systems are explicitly parallel
and provide very lightweight mechanisms for parallelism. Task-based systems for HPC tend to be next most efficient, and
provide additional features such as automatic dependency discovery and
data movement. Higher overhead systems tend to be designed primarily for
large-scale data analysis or workflows. It is worth
remembering that these are \emph{minimum}
effective task granularities. Applications with an average
task granularity of \emph{at least} this value can usually be expected
to execute efficiently. Typical task granularities will
generally be determined by the application domain being
considered. Most notably, for large-scale data analytics workloads, the higher METG values observed for Spark are sufficient. In contrast, for high-performance
scientific simulations, task granularities in the millisecond range
are useful, as such applications communicate (e.g., for halo
exchanges) much more frequently.

The least complicated pattern (stencil) is most favorable
to MPI, as it provides no
opportunity for task parallelism. The
dominating factor in this case is the overhead of executing a task, which is
minimal for MPI as the code simply executes tasks in alternation with
communication. The asynchrony
of other systems is pure overhead in this scenario.
MPI's advantage shrinks as complexity grows, and even reverses as task parallelism is
added in the form of multiple task graphs.

We omit Spark and Swift/T
with more complicated dependencies, as their higher overheads require
excessive problem sizes (beyond what completes in 6
hours) to reach 50\% efficiency.

\subsection{Scalability}
\label{subsec:scalability}

METG summarizes system overheads in a single number. This makes it possible to
evaluate how communication topology and latency impact METG at
different node counts, as shown in Figure~\ref{fig:metg-compute}.
We find that
systems with the smallest METG on one node have roughly an order
of magnitude higher METG at 256 nodes. Increased communication latencies require significantly larger tasks to
achieve the same level of efficiency, so apparent differences in
overhead at small node counts can matter much less or not at
all at larger node counts.

Most systems for HPC are highly scalable,
but this is not true of all the systems included in this
evaluation. Lower is better in Figure~\ref{fig:metg-compute}, and
flat is ideal. Lines that rise with node count indicate less than
ideal scaling. Most notably, Spark is primarily
intended for industrial data center applications with task
granularities measured in seconds. Spark uses a centralized
controller, which limits throughput, and this is visible in the figure
as the line for Spark immediately rises with node count. Keep in mind
that Spark is being evaluated here with a nontrivial dependence
pattern that is relatively unrepresentative of Spark's normal use
cases. Spark is more efficient with trivial parallelism, as described
in Section~\ref{subsec:number-of-dependencies}.

\begin{figure}[t]
\centering
\includegraphics[width=\columnwidth]{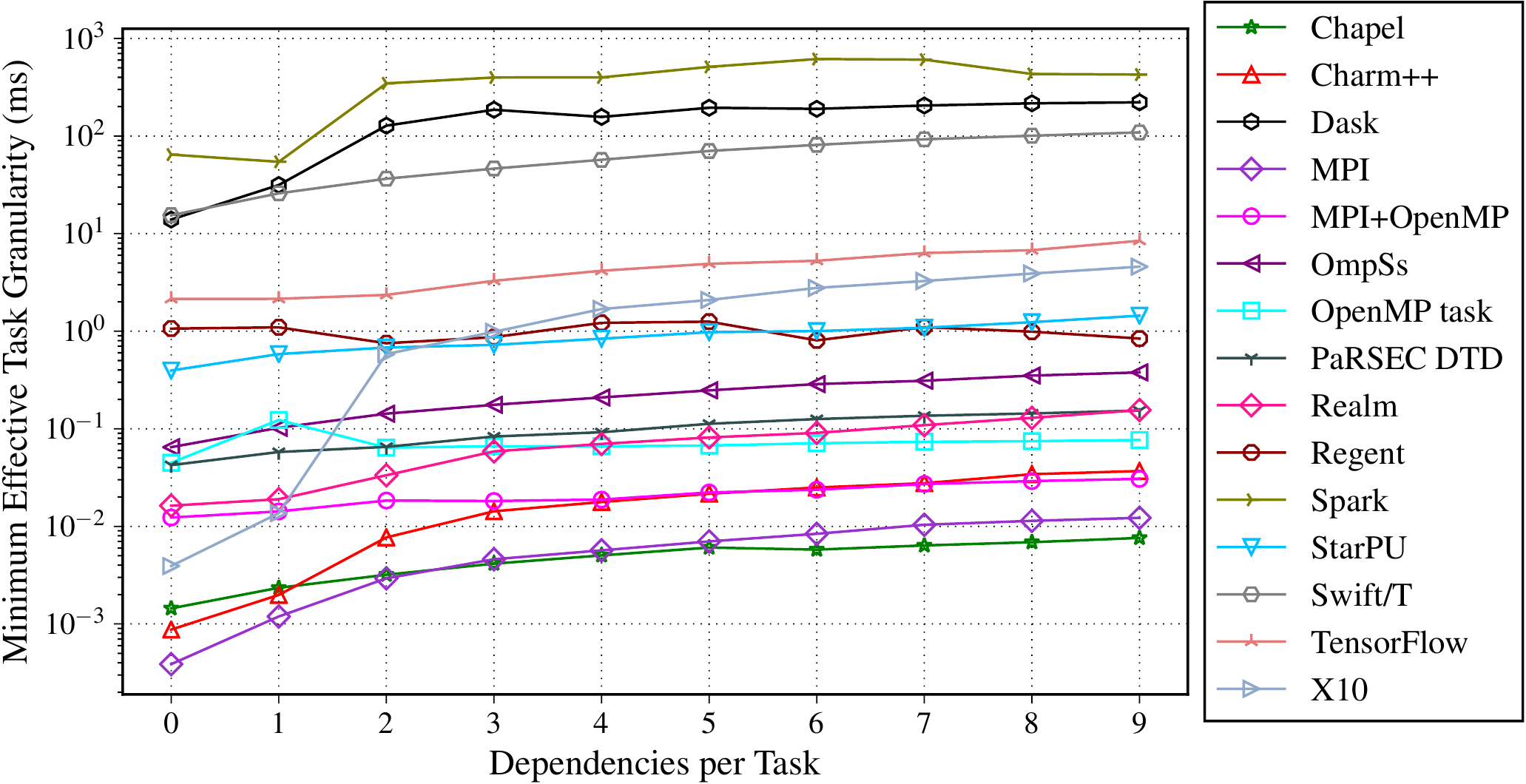}
\vspace{-0.6cm}
\caption{METG vs deps/task (nearest, 1 node). Lower is better.\label{fig:radix}}
\vspace{-0.25cm}
\end{figure}
 
\begin{figure*}[t!]

\subfloat[16 bytes per task dependency.\label{fig:efficiency-communication-16}]{
\includegraphics[width=\columnwidth]{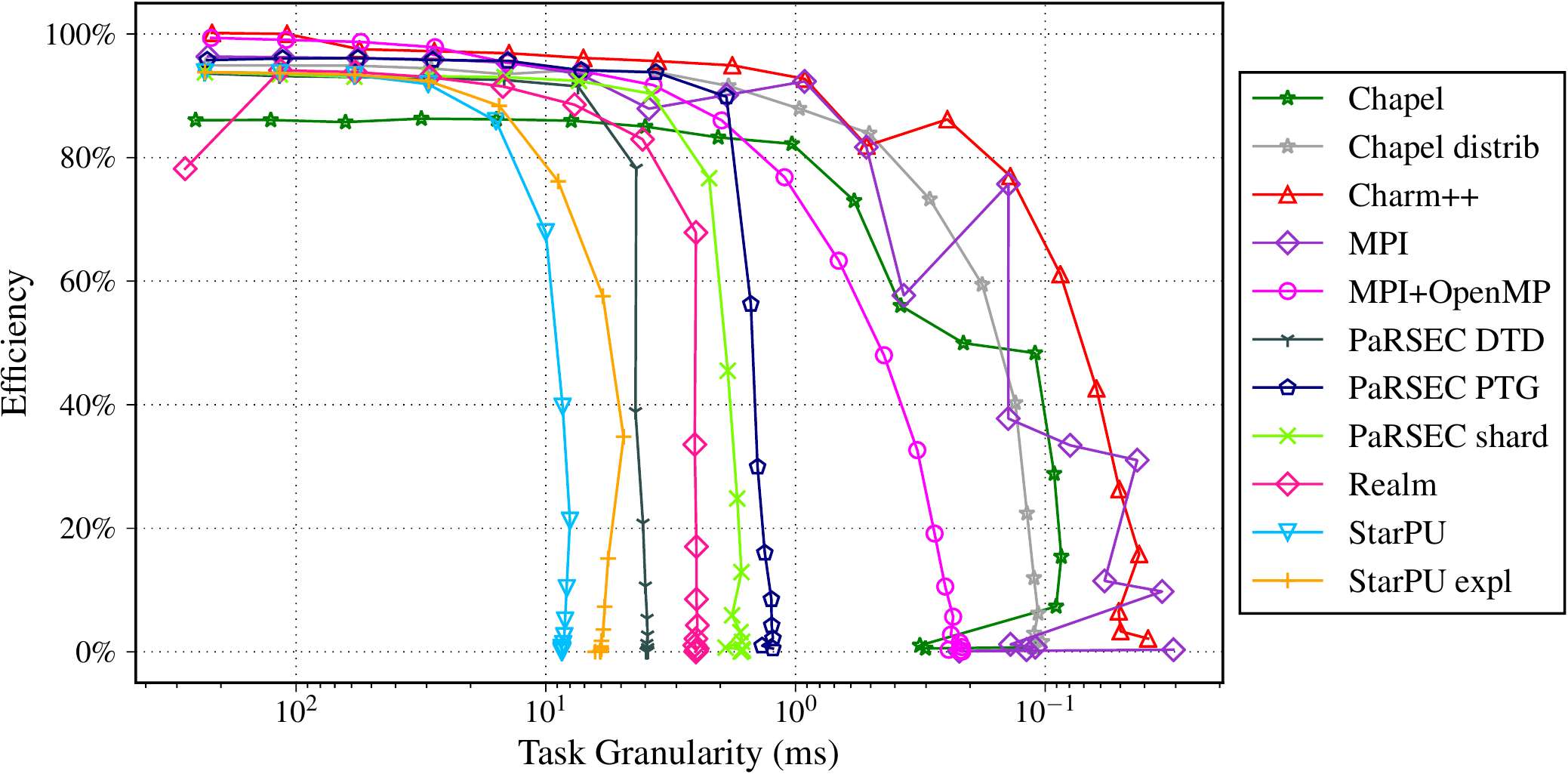}
}
\subfloat[65536 bytes per task dependency.\label{fig:efficiency-communication-65536}]{
\includegraphics[width=\columnwidth]{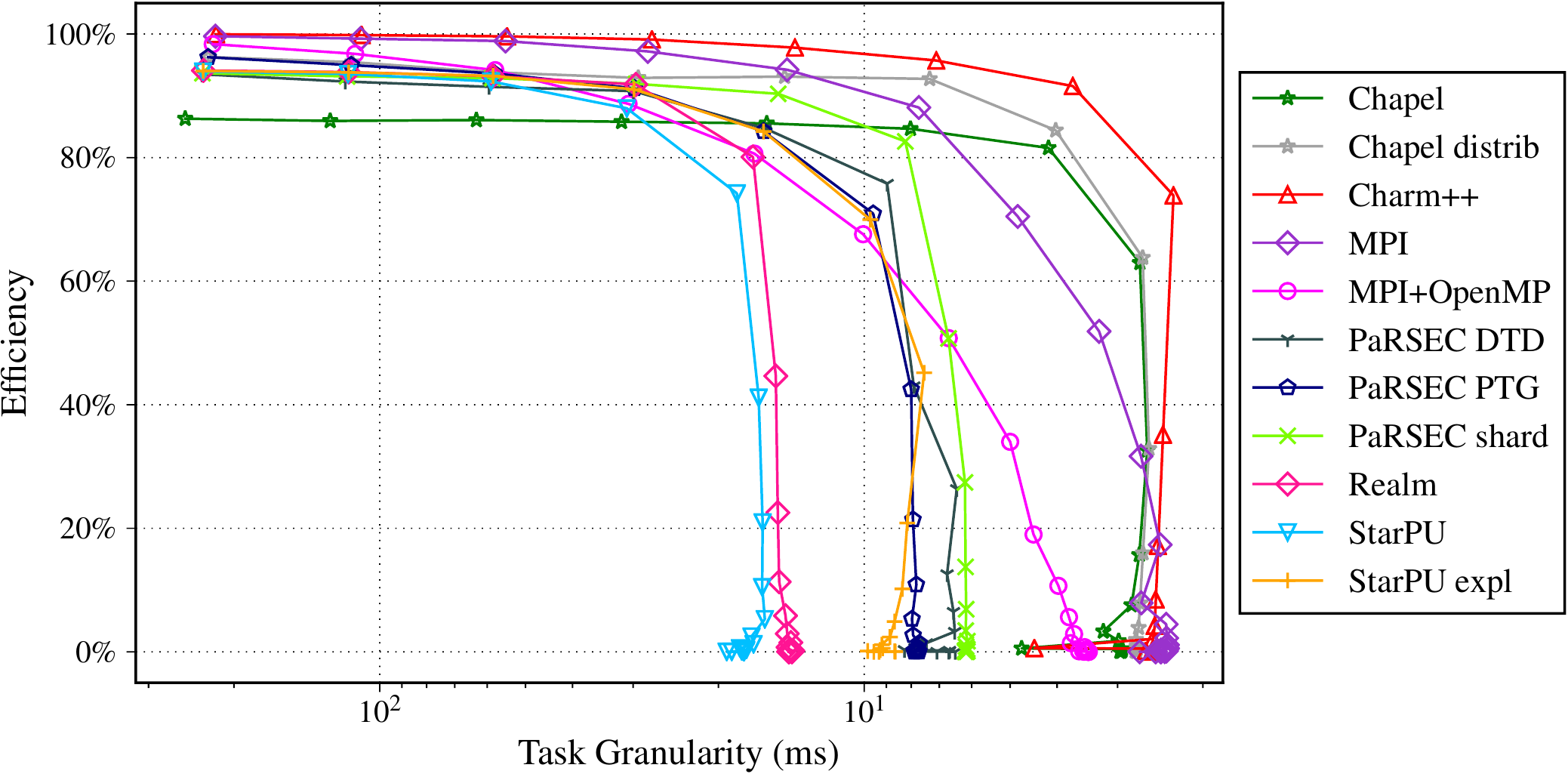}
}

\vspace{-0.15cm}
\caption{Efficiency vs task granularity for varying communication (spread, 5 deps/task, 4 independent graphs, 64 nodes). Higher is better.\label{fig:efficiency-communication}}
\vspace{-0.25cm}
\end{figure*}
 
PaRSEC, StarPU and Regent rely on runtime analysis that can suffer
from scalability bottlenecks if every node must consider the tasks
executing on all other nodes. Although all three systems offer ways to
improve scalability, these methods are not equally effective. PaRSEC
DTD and StarPU allow users to manually prune tasks to reduce overhead;
however the checks are not reduced to zero. Similarly, PaRSEC PTG uses
compile-time optimizations to avoid the need for manual pruning, but
still targets the PaRSEC DTD runtime and thus incurs some
overhead. Figure~\ref{fig:metg-compute} shows that these models do not
provide ideal scalability, as seen by METG values that rise with
increasing node count. Regent uses a compile-time optimization to
generate code with a constant overhead per
node~\cite{ControlReplication17}. Of the three systems, Regent is the
only one that achieves ideal scalability while preserving its
original, implicitly parallel programming model. The others can
achieve ideal scalability but require increasing levels of manual
intervention. PaRSEC \lstinline{shard} includes additional manual
optimizations over DTD to completely eliminate dynamic checks. StarPU
\lstinline{expl} is written in an explicitly parallel style using MPI
for communication, and thus avoids any analysis bottleneck. These
results indicate that the underlying systems are capable of scalable
execution, but that the dynamic checks incurred by the implicitly
parallel programming models hinder that scalability.

\subsection{Number of Dependencies}
\label{subsec:number-of-dependencies}

The number of dependencies per task has a strong influence on
overhead, as shown in
Figure~\ref{fig:radix}. This plot shows METG(50\%) for the nearest
dependence pattern, when varying the number of dependencies per task
from 0 to 9.

The ratio in METG between 0 and 3 dependencies per task ranges from
$0.82\times$ to $250\times$ (mean $21\times$, std. dev. $64$). The
large standard deviation shows that the sensitivity of system overhead
to the dependency pattern varies widely. The largest ratios are among
systems that perform runtime work inline. For example, MPI achieves an
METG of 390 ns with 0 dependencies, but this rises to 4.6 \textmu{}s
with 3 dependencies, a $12\times$ increase. This is unsurprising, as
with 0 dependencies no \lstinline[language=C++]{MPI_Isend} calls are
issued at all. Clearly, choosing a representative dependence pattern
is important when estimating the performance of a workload or class of
workloads.

\subsection{Overlapping Communication and Computation}

Also of interest is the ability to
hide communication latency in the presence of task
parallelism. Figure~\ref{fig:efficiency-communication} plots efficiency with varying amounts of
communication, determined by the number
of bytes produced by each task (and therefore communicated with each
task dependency).

Asynchronous systems such as Charm++ demonstrate two benefits in
these plots. First, by overlapping communication with computation,
such systems execute smaller task granularities at higher
levels of efficiency compared to the MPI
implementations. Second, the asynchrony and scheduling flexibility from
executing multiple graphs also makes the curves smoother,
as spikes in latency due to interference from other jobs can be
mitigated, leading to more predictable performance, especially at
smaller message sizes.

The effectiveness of such overlap can be influenced by the scheduling
policies of the underlying system. For example, Chapel's default
scheduler uses a round-robin policy; we see in
Figure~\ref{fig:efficiency-communication} that this approach fails to
take full advantage of the available task parallelism. A work-stealing
scheduler (Chapel \lstinline{distrib}) is able to recover this
performance.

\subsection{Load Imbalance}

\begin{figure}[t]
\centering
\includegraphics[width=\columnwidth]{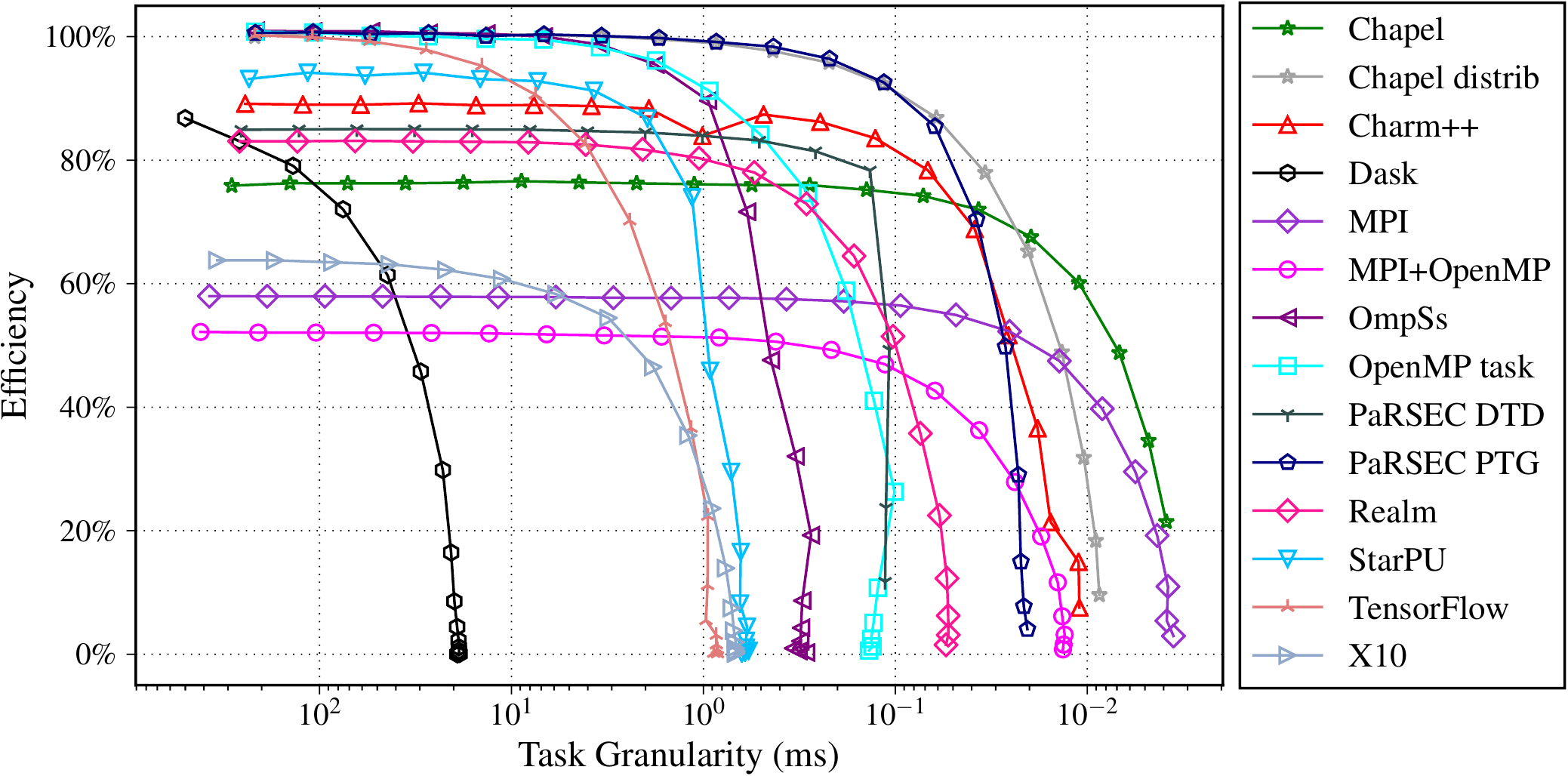}

\vspace{-0.15cm}
\caption{Efficiency vs task granularity under load imbalance (nearest pattern, 5 deps/task, 4 task graphs, 1 node). Higher is better.\label{fig:efficiency-imbalance}}
\vspace{-0.1cm}
\end{figure}
 
One advantage of asynchronous execution is
the ability to mitigate load imbalance with little or no additional programmer effort, especially in the presence of
task parallelism. To quantify this effect,
Figure~\ref{fig:efficiency-imbalance} plots task granularity vs.
efficiency under load imbalance where each task's duration is multiplied by a uniform random variable in [0,~1). Task durations are generated with a deterministic
pseudo random number generator with a consistent seed to ensure
identical durations for all systems.

The MPI Task Bench, with its distinct computation and communication phases,
suffers the most under load imbalance. The biggest
difference is at large task granularities, where the imbalance
effectively puts an upper bound on efficiency. At smaller task
granularities the effect shrinks and may even reverse as systems hit
their fundamental limits due to overhead.

The remaining differences are due primarily to
different scheduling behaviors. The execution of 4
simultaneous task graphs only partially mitigates the
load imbalance between tasks. Systems that provide an
additional on-node work stealing capability (such as Chapel with the \texttt{distrib} scheduler) see additional gains in
efficiency at large task granularities. However, the use of
work-stealing queues can also impact throughput at small task
granularities. For example, Chapel's default (non-work-stealing) scheduler outperforms \texttt{distrib} at very small task granularities. We do not consider Charm++ load balancers because the imbalance is \emph{non-persistent} (i.e., timestep $t$ is uncorrelated with timestep $t+1$). We leave analysis of persistent load imbalance to future work.

\subsection{Heterogeneous Processors}

\begin{figure}[t]
\centering
\includegraphics[width=\columnwidth]{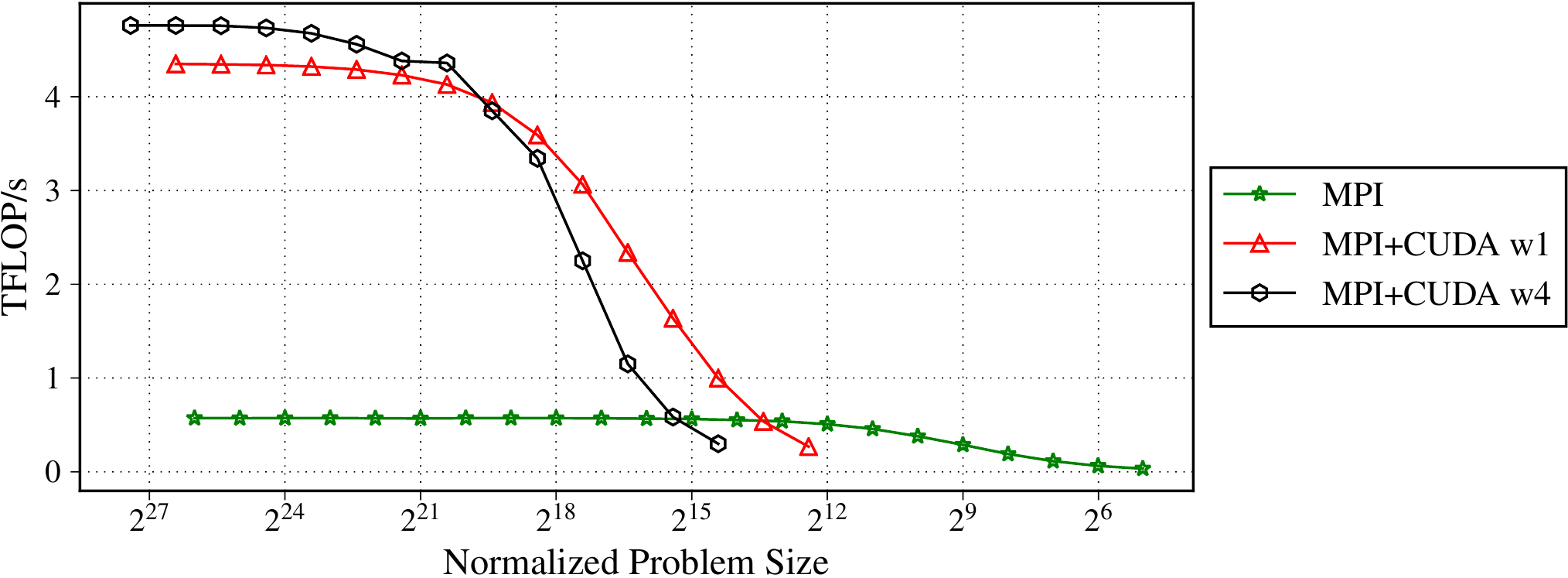}
\vspace{-0.5cm}
\caption{GPU FLOP/s vs normalized problem size (stencil, 1 node). Higher is better.\label{fig:cuda-efficiency}}
\vspace{-0.25cm}
\end{figure}
 
To determine the cost of scheduling tasks on GPUs, Figure~\ref{fig:cuda-efficiency} compares MPI
and MPI+CUDA on Piz Daint~\cite{PizDaint}. The CUDA
compute kernel achieves $4.759 \times 10^{12}$
FLOP/s, which is very close to the officially reported number
$4.761 \times 10^{12}$. The CPU achieves $5.726 \times 10^{11}$
FLOP/s. Note that the kernels perform different numbers of
operations as the GPU requires more work to reach peak
performance. The x-axis in Figure~\ref{fig:cuda-efficiency} is
normalized to keep FLOPs constant for a given problem
size.

Our MPI+CUDA code uses an offload model with data copied to/from the GPU every step. In our tests, \lstinline{w1} uses 1 task per GPU, whereas
\lstinline{w4} overdecomposes, using 4 MPI
ranks per GPU to push work to the GPU in parallel.
\lstinline{w4} achieves higher FLOP/s but
drops more rapidly at small problem sizes, due to the overhead of
running $4\times$ as many CUDA kernels. Either way,
GPUs require more work to achieve high performance, and the overhead
of copying data dominates at small task
granularities, where CPUs achieve higher
performance. While Figure~\ref{fig:cuda-efficiency} is not couched in
terms of METG (as peak performance on CPU and GPU are very
different), the conclusion here is similar to
Section~\ref{subsec:scalability}: the cost of sending data and
tasks to GPUs imposes a floor on task granularity relative to CPUs, reducing the advantage at small task granularities of
very lightweight mechanisms such as those in MPI.

\subsection{Validating METG with Time to Solution}

We can use METG to predict the scalability of a code. Figure
\ref{fig:strong-scaling} shows strong scaling for three Task Bench
implementations with the stencil pattern. Lines marked ``actual''
represent strong scaling measurements, while ones marked ``limit 50\%'' are
computed by multiplying METG(50\%) by tasks per core (in this case,
1000) to obtain wall clock time. Intuitively, ``limit'' is the smallest time to solution that
can be achieved for \emph{any} problem size at that node count, while
maintaining 50\% efficiency. Data points where ``actual'' falls below
``limit'' mark points where strong scaling parallel efficiency is less
than 50\%.

There are two points of interest on each ``limit 50\%'' curve: the point
where it intersects ``actual'' and the point where it intersects an
ideal scaling curve, computed by taking the initial time to solution
and assuming linear scaling. These points are marked for Charm++ in
Figure~\ref{fig:strong-scaling} with a black square and black circle,
respectively.
Notably, the ideal-limit intersection (black circle) requires only a
run of the application on one node, combined with METG(50\%)
measurements, to estimate the strong scalability of
the code:
i.e., the smallest time to solution,
and the number of nodes at which that time is achieved, while
maintaining at least 50\% efficiency. In
Figure~\ref{fig:strong-scaling}, the error in the estimate is the distance
between the black square and circle; these are separated by a factor of
$1.22\times$ in node count, and $1.27\times$ in time to solution.

\begin{figure}[t]
\centering
\includegraphics[width=\columnwidth]{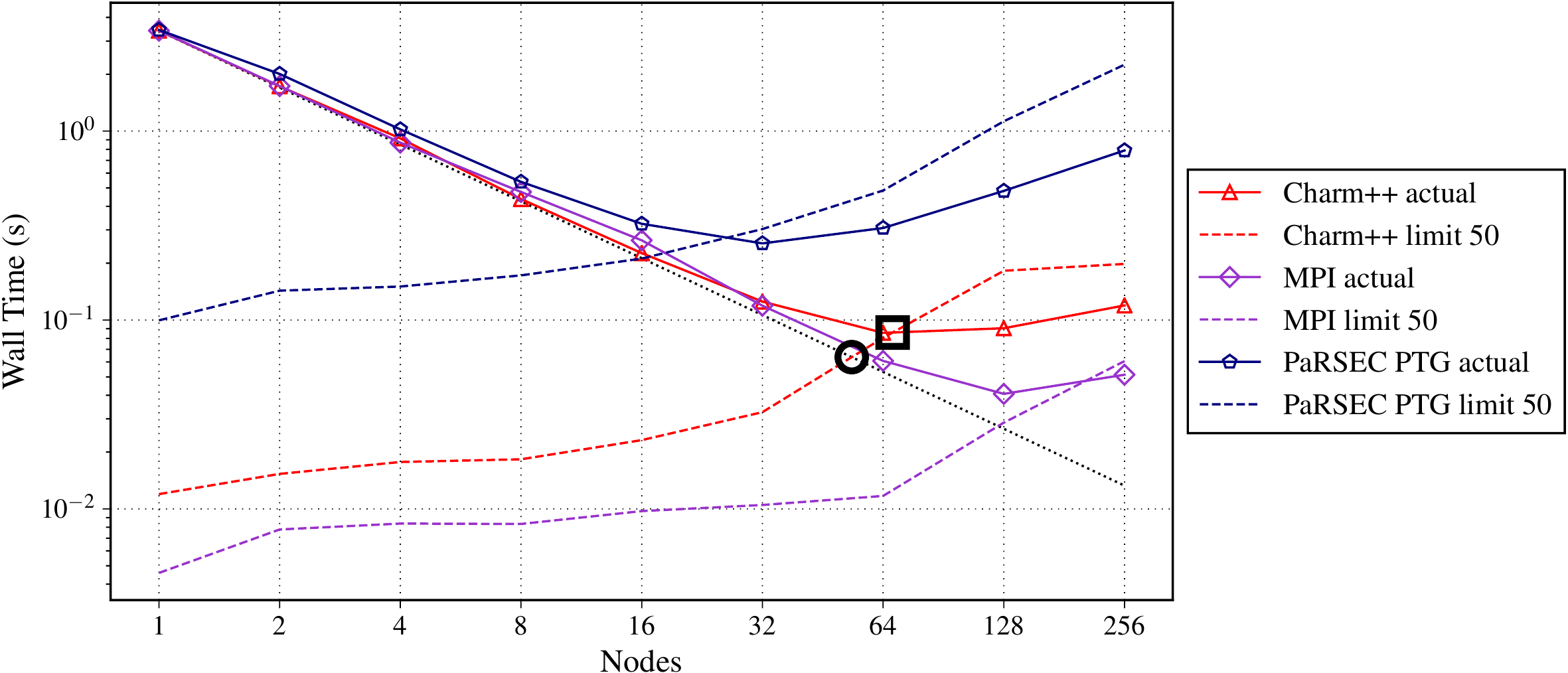}
\vspace{-0.6cm}
\caption{Strong scaling vs efficiency-limited time to solution with problem size $2^{20}$ (stencil). Lower is better.\label{fig:strong-scaling}}
\vspace{-0.35cm}
\end{figure}
 \begin{table}[t]
\centering
\small
\begin{tabular}{r | c | c | c | c}
        & \multicolumn{2}{|c|}{Nodes} & \multicolumn{2}{|c}{Time to Solution} \\
Pattern & mean & std.~dev. & mean & std.~dev. \\
\hline
stencil & $1.64\times$ & 0.278 & $1.26\times$ & 0.246 \\
nearest & $1.75\times$ & 0.268 & $1.15\times$ & 0.331 \\
spread  & $1.43\times$ & 0.288 & $1.29\times$ & 0.206 \\
nearest, 4 graphs & $1.96\times$ & 0.680 & $1.26\times$ & 0.702
\end{tabular}

\vspace{-0.20cm}
\caption{Factor of separation between limit-ideal and limit-actual intersections across all 12 programming systems tested in Section~\ref{subsec:scalability}.\label{tab:metg-predict-strong}}
\vspace{-0.5cm}
\end{table}
 
Table~\ref{tab:metg-predict-strong} expands this comparison to the 4
patterns tested in Section~\ref{subsec:scalability} for all 12
programming systems that scale well enough to evaluate the separation
between the limit-ideal and limit-actual intersections. We see that
overall, the mean separation is at most $1.96\times$ in node count
and at most $1.29\times$ in time to solution, making this a
useful way to predict strong scaling in the 4 patterns we tested in
Section~\ref{subsec:scalability}.

\zap{
\begin{figure}[t]
\centering
\includegraphics[width=\columnwidth]{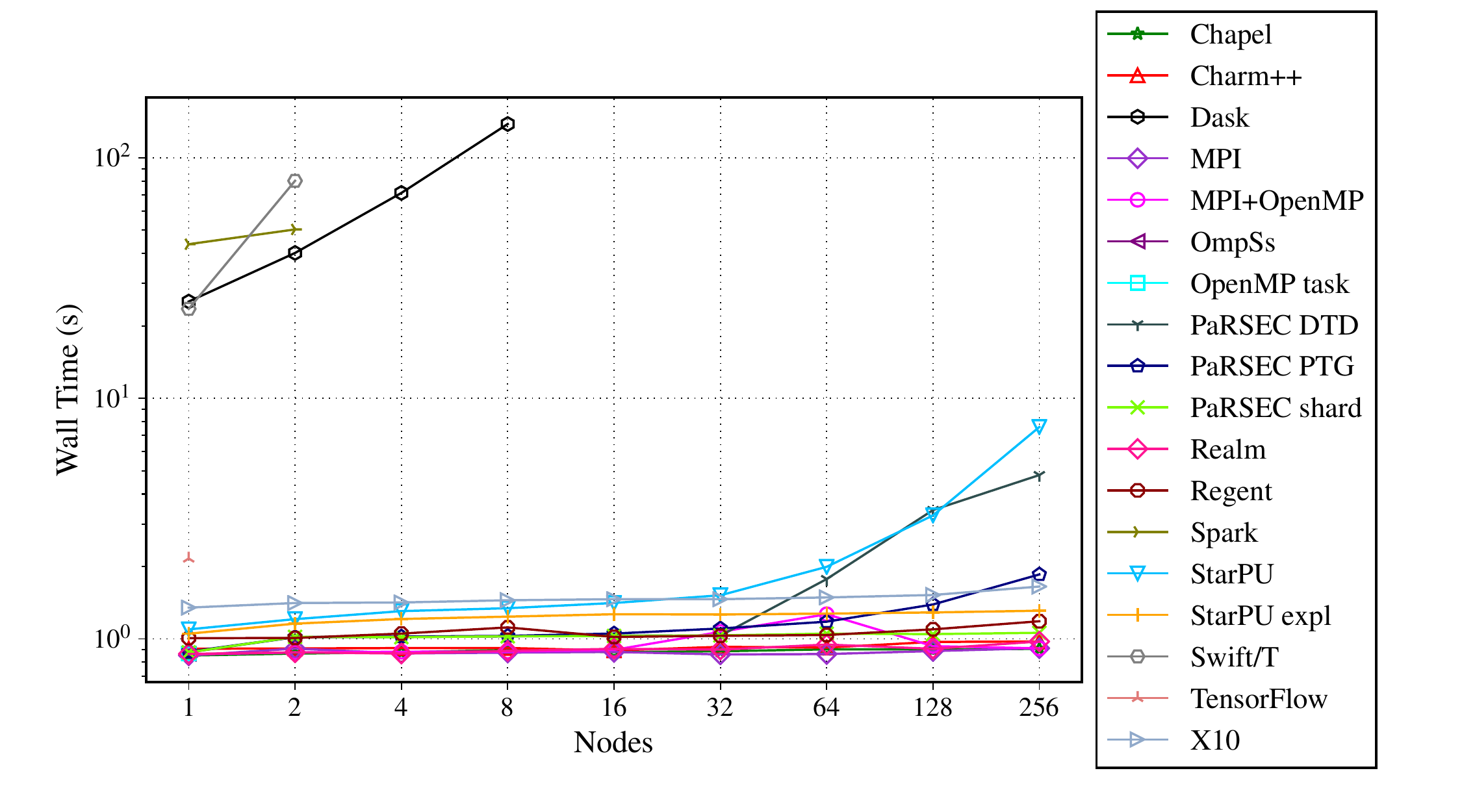}
\vspace{-0.5cm}
\caption{Weak scaling with problem size $2^{18}$ per node (stencil). Lower is better.\label{fig:weak-scaling}}
\vspace{-0.05cm}
\end{figure}
 \begin{figure}[t]
\centering
\includegraphics[width=\columnwidth]{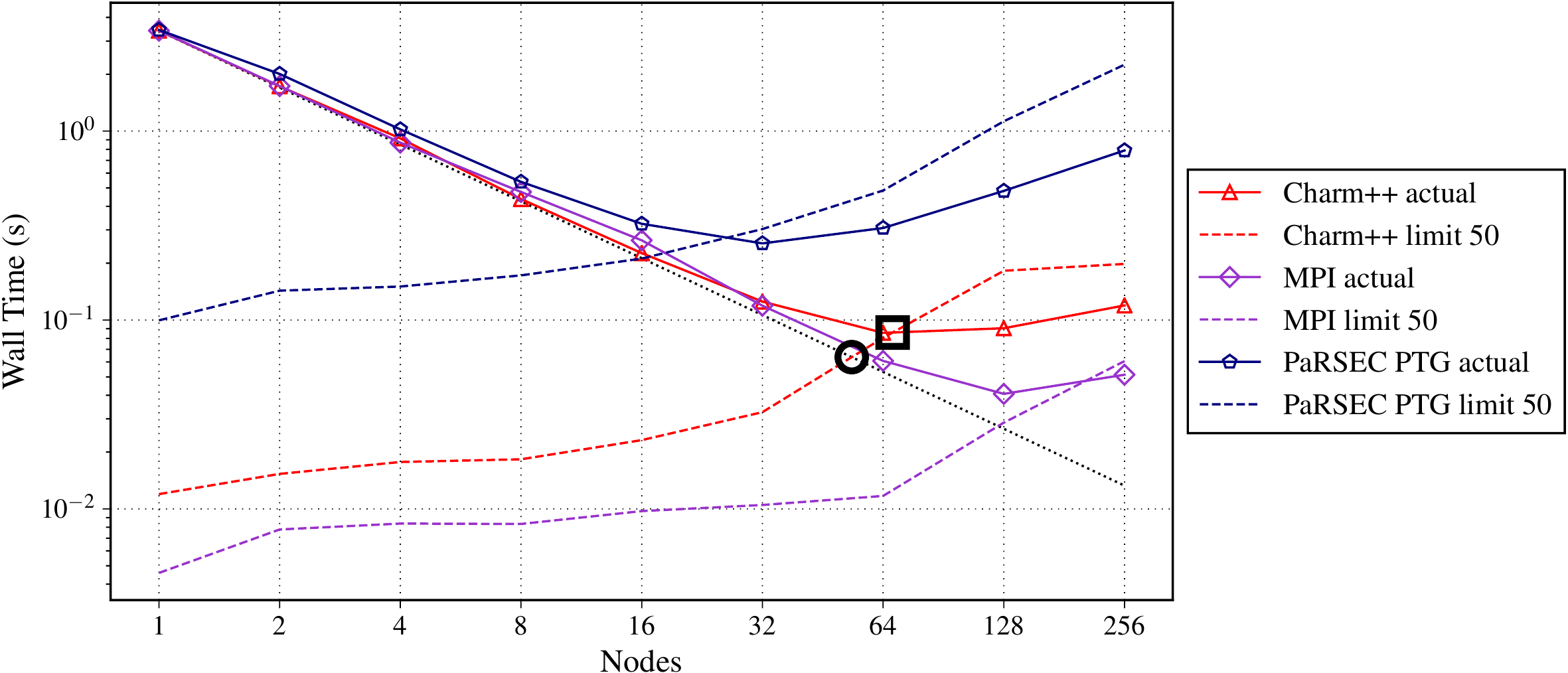}
\vspace{-0.6cm}
\caption{Strong scaling vs efficiency-limited time to solution with problem size $2^{20}$ (stencil). Lower is better.\label{fig:strong-scaling}}
\vspace{-0.35cm}
\end{figure}
 \subsection{Comparison with Weak and Strong Scaling}

Figures~\ref{fig:weak-scaling} and \ref{fig:strong-scaling} show
conventional strong and weak scaling of the stencil pattern for
comparison with the METG results in
Figure~\ref{fig:metg-compute-stencil}. At small node counts, where the
average task granularity is larger than the METG for most systems, the
performance is ideal. For several systems METG rises with node count,
and this becomes visible in the weak and strong scaling graphs as
non-ideal performance. Thus the shape of the weak and strong scaling
graphs effectively interpolates between ideal performance and the METG
curve depending on how close each system is to being limited by
overhead.
}  \section{Performance Issues Discovered}
\label{sec:case-study}

Task Bench is useful not only as a tool for evaluating programming
system performance, but also for discovering potential areas for
improvement. During the development of Task Bench, we identified a
number of performance issues in the underlying programming
systems. These discoveries were possible because of the flexibility of
Task Bench, and our ability to rapidly run new experiments with a
variety of application scenarios. All issues were reported to and
acknowledged by the respective system's developers, and are either
fixed or worked around in our experiments. We describe them below.

Realm and Chapel both use DMA subsystems optimized for large
copies. Early Task Bench experiments revealed high METG values,
diagnosed by the Realm/Chapel developers as overhead due to the
cost of scheduling small copies. Subsequent improvements in Realm and
Chapel improved small copy overheads (and thus METG) by over an order
of magnitude in the case of Realm, and by $2\times$ in the case of
Chapel. These improvements affect any application where fine-grained
data movement is needed, which is particularly relevant in strong scaling
regimes when running on large numbers of nodes.

Further analysis of Realm efficiency indicated many overheads due to
dynamic graph construction. The Realm developers implemented a new
``subgraph'' API in response to this feedback to amortize the cost of
repeatedly constructing isomorphic task graphs. This API is used in
Task Bench to achieve further speedups in the Realm implementation.

Chapel uses a naive round-robin scheduler by default, which can lead
to unexpected load balance issues because certain language features
(such as remote array assignment) implicitly generate tasks. This
resulted in poor peak performance, even at large task granularities,
which in some cases made it impossible to measure METG (because peak
efficiency did not exceed 50\%). Chapel's other schedulers add
overhead, resulting in higher METGs. Based on guidance from the Chapel team, we worked around this with
a \lstinline{serial} block.

PaRSEC uses task pruning to reduce analysis work performed on each
node, and thus improve scalability at large node
counts. Initial Task Bench results achieved less than the expected
scalability: METG was rising too quickly with node count.  The PaRSEC
developers diagnosed a bug in task pruning---the optimization was failing to trigger---and it was
subsequently fixed. We also implemented a version of the PaRSEC code,
\lstinline{shard}, which uses manual pruning to demonstrate that
additional gains may still be possible.

Early Task Bench results for Dask revealed that the cost of scheduling
a task was $\mathcal{O}(N)$ where $N$ is the number of tasks in a task
graph, causing overall cost for a task graph with $N$ nodes to be
$\mathcal{O}(N^2)$. This issue was reported to and confirmed by the
Dask developers; the bug is suspected to be in optimizations performed
on the task graph by Dask. As a workaround, our results in the paper use a
lower-level interface which does not suffer from this asymptotic
slowdown.

TensorFlow performs optimizations on task graphs prior to execution,
including a constant folding pass. The Task Bench implementation uses
aggressive loop unrolling to minimize overheads, resulting in the task
graph being marked as constant. Because TensorFlow optimizations are
performed on a single core, this resulted in sequential execution of
the entire graph, and Task Bench experiments timed out. As a
workaround, additional inputs are fed to nodes in the graph to make
them non-constant, forcing them to be executed by TensorFlow's parallel
scheduler. Though this was exacerbated by Task Bench's approach to
unrolling loops, it affects any task graph with constants that are
expensive to compute where the users may wish to perform
constant-folding in parallel.

\zap{
Initial experiments with TensorFlow revealed that the Task
Bench implementation was not able to use all cores on a node (making it impossible to measure METG since efficiency was below
50\%). The TensorFlow developers diagnosed this as an issue in
constant folding. The entire Task Bench task graph was being detected
as constant, and the constant-folding pass runs on one core. As a
workaround, the developers suggested marking the task graph inputs as
non-constant so that TensorFlow's standard task scheduler could be
used.
}

In all cases, we found bugs that are applicable outside of Task Bench
and that impact metrics other than METG. Notably, the scalability
issues and asymptotic complexity have a growing impact with larger
node counts and will eventually become visible with nearly any
application. In other cases, Task Bench and METG made it possible to
identify performance degradation at the extremes of application
configurations which might remain hidden with full-size applications,
particularly when evaluated with weak and strong scaling alone. The
improvements motivated by our findings are nonetheless relevant in
strong-scaling regimes in a variety of applications, particularly as
node and core counts grow.

\zap{
We found further areas for improvement in schedulers (Realm, Chapel
and Dask), task pruning for scalability (PaRSEC), and optimizations on
the task graph (TensorFlow). The impact of these optimizations ranged
from small constant factors to even asymptotic improvements in
performance and scalability (in the case of PaRSEC and Dask). All
users will benefit from the improvements driven by our discoveries.
}  \section{Related Work}
\label{sec:related-work}

\brokenpenalty=0

Parallel and distributed programming systems are often
evaluated using proxy- or mini-apps, or
microbenchmarks. Mini-apps are explicitly derived from larger
applications and hence have the advantage of bearing some
relationship to the original. This advantage typically does not hold
for microbenchmarks.

\brokenpenalty=\oldbrokenpenalty

Though smaller than full applications, mini-apps can be challenging to
implement to a level of quality sufficient for conducting comparative
studies between programming systems. The largest studies
we know of consider at most 7 and 6 programming systems,
respectively~\cite{LULESH13, Deakin19}, and the latter only considers
on-node programming models. In both cases, the mini-apps under study
require a separate, tuned implementation (in contrast to Task
Bench). Other studies usually lack a comprehensive evaluation, even if
multiple implementations are available:

\begin{itemize}
\item The PENNANT reference implementation supports
MPI/OpenMP/MPI+OpenMP~\cite{PENNANT}. A follow-up paper presents a
Regent implementation~\cite{Regent15}.

\item One follow-up paper for the mini-app CoMD describes a Chapel
implementation~\cite{CoMDChapel16} (comparison against reference
only). Additional follow-up papers consider aspects of the reference
implementation only~\cite{CoMDLoadImbalance17,
  CoMDThreadedModels14}.

\item A report on the Mantevo project~\cite{Mantevo09} describes a number of
mini-apps, but only includes self-comparisons based on reference
implementations.

\item A report on MiniAero~\cite{SandiaReportManyTaskRuntimes15} describes
four implementations of the mini-app, but only includes performance
results for three, of which only two can be compared in an
apples-to-apples manner as the last implementation uses structured
rather than unstructured meshes. A follow-up describes another
implementation in Regent~\cite{Regent15} (comparison vs.~reference
only).
\end{itemize}

Microbenchmarks can be easier to implement, but do not address the
asymptotic costs of implementation. PRK Stencil~\cite{PRK14} contains a
2D stencil and is evaluated on implementations in MPI, SHMEM, UPC,
Charm++, and Grappa~\cite{PRKRuntimes16}. The NAS benchmark
suite~\cite{NAS91, NAS95} consists mostly of small kernels for dense
or sparse matrix computations and has implementations in
OpenMP~\cite{NASOpenMP99}, MPI and MPI+OpenMP~\cite{NASMPIOpenMP00},
and Charm++~\cite{NASCharm96}. PRK requires $\mathcal{O}(n)$ effort to
implement (for $n$ systems) by virtue of being only a single
computational pattern, while NAS requires $\mathcal{O}(mn)$ overall
effort (for $m$ patterns and $n$ systems). In contrast, Task Bench
requires $\mathcal{O}(m+n)$ effort and is easily extended to cover new
systems or patterns with $\mathcal{O}(1)$ effort for each additional
system or pattern.

System-specific benchmarks quantify specific aspects
of system performance, such as MPI communication or collective
latency~\cite{MPPTest99, MPIBench01}. These measurements typically do
not generalize beyond the immediate system they measure.

\textsc{coNCePTuaL}~\cite{Conceptual07} is a domain-specific language
for writing network performance tests. \textsc{coNCePTuaL} and Task
Bench both enable the easy creation of new benchmarks, though
\textsc{coNCePTuaL} does so via scripting whereas Task Bench provides
a set of configurable parameters. \textsc{coNCePTuaL} also targets a
lower level of abstraction, optimized more for testing messaging
layers, whereas Task Bench is closer to application level and
therefore enables comparisons of a broader set of parallel and
distributed programming systems.

Limit studies of task scheduling throughput in various runtime systems
often make additional assumptions. A popular assumption is the use of
trivially parallel tasks~\cite{Canary16, Armstrong14}, which as shown
in Section~\ref{subsec:number-of-dependencies} underestimates (often
substantially) the cost of scheduling a task and can also impact scalability.
 \section{Conclusion}
\label{sec:conclusion}

Task Bench is a new approach for evaluating the performance of
parallel and distributed programming systems. By separating the
specification of a benchmark from implementations in various
programming systems, Task Bench reduces overall developer effort to
$\mathcal{O}(m + n)$ (for $m$ benchmarks on $n$ systems) rather than
$\mathcal{O}(mn)$ as has been the case for all previous
benchmarks that we know of. This has enabled us to explore a broad space of
application scenarios and to do so with a large number of programming
systems. Our experiments have enabled the following
insights:

\begin{itemize}
\item Evaluations of programming system performance should avoid using
  TPS, or strong or weak scaling to characterize overheads, as these metrics do
  not constrain the useful work achieved. Instead a metric with
  constrained efficiency, such as METG(50\%), is needed to ensure that
  measurements are representative and fair.

\item METG for current distributed programming systems varies by over
  5 orders of magnitude.  Clearly understanding the needed task
  granularity is an important consideration in choosing a programming
  system for a new application.

\item While some systems support METG(50\%) as small as 390~ns, this applies only to trivial dependencies and small CPU-based clusters. A number of factors (nontrivial dependencies, accelerators and cluster sizes in the hundreds of nodes) raise
  the METGs that can be achieved by over an order of magnitude: 100~\textmu{}s is a reasonable bound for most applications running at scale with current technologies.

\item Systems that support asynchronous execution show benefits under
  balanced computation
  and communication, and load imbalance. However, these gains can be nullified by
  high baseline overheads.

\item Task-based systems that rely on runtime analysis for the
  discovery of parallelism can suffer from sequential bottlenecks that
  limit scaling. Existing, dynamic task pruning techniques are not
  sufficient to fully mitigate this bottleneck, while static,
  compile-time approaches are able to do so.

\item Systems for large scale data analysis require very large tasks
  (tens of seconds) to scale beyond small node counts,
  reflecting the very coarse tasks and lack of need for strong scaling
  in current workloads.

\item Task Bench has proven effective in finding performance issues
  and has lead to substantial improvements in several systems
  we study.
\end{itemize}

\zap{
We have implemented Task Bench for a broad set of programming systems
spanning large scale data analytics and HPC. However, there are
systems not represented in our evaluation that would be interesting to
consider in the future. These include GASNet~\cite{GASNET07},
Habanero~\cite{Habanero11}, Hadoop~\cite{Hadoop},
HPX~\cite{Kaiser2014}, Nimbus~\cite{Nimbus17}, OCR~\cite{OCR14},
OpenSHMEM~\cite{OpenSHMEM10}, Ray~\cite{Ray18}, and UPC~\cite{UPC99}.
} 

Not considered in our analysis is the impact of programming system
features on programmer productivity and performance portability. Most
applications do not operate at the absolute extreme of runtime-limited
performance, and thus may choose to trade overhead for
usability. Our study helps to quantify the performance side of that
tradeoff so that users can be better informed and developers can see
the impact that features have on the performance of their programming
systems.

\section*{Acknowledgment}

This material is based upon work supported by the U.S. Department of
Energy, Office of Science, Office of ASCR, under the contract number
DE-AC02-76SF00515, by National Science Foundation under Grant
No. ACI-1450300, and the Exascale Computing Project (17-SC-20-SC), a
collaborative effort of the U.S. Department of Energy Office of
Science and the National Nuclear Security Administration, under prime
contract DE-AC05-00OR22725, and UT Battelle subawards 4000151974 and
89233218CNA000001. Experiments on the Cori supercomputer were
supported by the National Energy Research Scientific Computing Center,
a DOE Office of Science User Facility supported by the Office of
Science of the U.S. Department of Energy under Contract
No. DE-AC02-05CH11231, and experiments on Piz Daint were supported by
the Swiss National Supercomputing Centre (CSCS) under project ID
d80. A special thanks to thank Katie Antypas for her support.

The authors would like to thank the developers of the programming
systems tested in this paper. Their support enabled us to produce
high-quality implementations of the systems under study. The
developers consulted include (alphabetically by system): Chapel:
Bradford L.~Chamberlain and Elliot Ronaghan; Charm++: Laxmikant Kalé
and Sam White; Dask: Matthew Rocklin; MPI: Samuel K.~Gutiérrez and Wei
Wu; OmpSs: Víctor López and Vicenç Beltran Querol; OpenMP: Alejandro
Duran and Jeff R.~Hammond; PaRSEC: George Bosilca and Qinglei Cao;
Realm: Seema Mirchandaney and Sean Treichler; Regent: Michael Bauer,
Wonchan Lee, Seema Mirchandaney and Elliott Slaughter; Spark: Matei
Zaharia; StarPU: Samuel Thibault; Swift/T: Justin M.~Wozniak;
TensorFlow: Jing Dong, Peter Hawkins and Mingsheng Hong; X10: David
Grove, Sara S.~Hamouda and Josh Milthorpe.
 
\bibstyle{IEEEtran}
\bibliography{bibliography}

\end{document}